\documentclass{aa}  
\usepackage{graphicx}
\usepackage[dvipsnames]{xcolor}

\usepackage{txfonts}
\usepackage{natbib}
\usepackage{xcolor}
\usepackage{url}
\usepackage{multicol}
\usepackage{multirow}
\usepackage{fontawesome5}
\usepackage{hyperref}
\usepackage[utf8]{inputenc}
\usepackage{amsmath}
\usepackage{nccmath}
\usepackage{amsfonts}
\usepackage{amssymb}
\usepackage{graphicx}
\usepackage{comment}
\usepackage{caption}
\usepackage{subcaption}
\usepackage{soul}
\usepackage[normalem]{ulem}
\usepackage{lscape}
\usepackage{longtable}
\usepackage{gensymb}

\newcommand\soutblue{\bgroup\markoverwith{\textcolor{blue}{\rule[.5ex]{2pt}{1.5pt}}}\ULon}

\newcommand\Tstrut{\rule{0pt}{2.6ex}}         
\newcommand{\sr}{\rule[-0.2cm]{0pt}{0.4cm}}
\newcommand{\srb}{\rule[-0.25cm]{0pt}{0.5cm}}

\definecolor{ForestGreen}{rgb}{0.13, 0.55, 0.13}
\definecolor{Orchid}{rgb}{0.85, 0.44, 0.84}
\usepackage{color}

\begin{document}

\title{TDCOSMO. VII. Boxyness/discyness in lensing galaxies:  Detectability and impact on $H_0$.}

\author{Lyne Van de Vyvere\inst{1}\fnmsep \thanks{\email{lyne.vandevyvere@uliege.be}}
          \and
          Matthew R. Gomer \inst{1}
          \and
          Dominique Sluse\inst{1}
          \and
          Dandan Xu \inst{2}
          \and
          Simon Birrer \inst{3,4}
          \and
          Aymeric Galan \inst{5}
          \and
          Georgios Vernardos \inst{5}
          }

\institute{STAR Institute, Quartier Agora - All\'ee du six Ao\^ut, 19c B-4000 Li\`ege, Belgium \and Department of Astronomy, Tsinghua University, Beijing, 100084, China \and Kavli Institute for Particle Astrophysics and Cosmology and Department of Physics, Stanford University, Stanford, CA 94305, USA \and SLAC National Accelerator Laboratory, Menlo Park, CA, 94025 \and Institute of Physics, Laboratory of Astrophysics, Ecole Polytechnique F\'ed\'erale de Lausanne (EPFL), Observatoire de Sauverny, 1290 Versoix, Switzerland}

   \date{Received 15 June 2021 / Accepted 1 December 2021 }

  \abstract
   {In the context of gravitational lensing, the density profile of lensing galaxies is often considered to be perfectly elliptical. Potential angular structures are generally ignored, except to explain flux ratios of point-like sources (i.e. flux ratio anomalies). Surprisingly, the impact of azimuthal structures on extended images of the source has not been characterised, nor has its impact on the $H_0$ inference. We address this task by creating mock images of a point source embedded in an extended source and lensed by an elliptical galaxy on which multipolar components are added to emulate boxy or discy isodensity contours. Modelling such images with a density profile free of angular structure allows us to explore the detectability of image deformation induced by the multipoles in the residual frame. 
   Multipole deformations are almost always detectable for our highest signal-to-noise ratio (S/N) mock data. However, the detectability depends on the lens ellipticity and Einstein radius, on the S/N of the data, and on the specific lens modelling strategy. Multipoles also introduce small changes to the time-delays. We therefore quantify how undetected multipoles would impact $H_0$ inference. When no multipoles are detected in the residuals, the impact on $H_0$ for a given lens is in general less than a few $\text{km}\,\text{s}^{-1}\,\text{Mpc}^{-1}$, but in the worst-case scenario, combining low S/N in the ring and large intrinsic boxyness or discyness, the bias on $H_0$ can reach 10-12 $\text{km}\,\text{s}^{-1}\,\text{Mpc}^{-1}$. If we now look at the inference on $H_0$ from a population of lensing galaxies with a distribution of multipoles representative of what is found in the light profile of elliptical galaxies, we find a systematic bias on $H_0$ of less than 1 \%.
   A comparison of our mock systems to the state-of-the-art time-delay lens sample studied by the H0LiCOW and TDCOSMO collaborations indicates that multipoles are currently unlikely to be a source of substantial systematic bias on the inferred value of $H_0$ from time-delay lenses.
   }
   
   
\maketitle

\section{Introduction}

The Hubble constant $H_0$ is a key parameter for understanding the past and future evolution of our Universe. Achieving a subpercent accuracy on its value is essential for exploring cosmological models beyond the standard $\Lambda$CDM. Several teams have achieved high-precision measurements using different methods \citep[e.g.][]{Abbott2018,Riess_cepheid,Freedman2020_tip_red,Pesce2020_megamaser,BAO_philcox2020,PlanckVI,Schombert2020_tullyfisher,HolicowXIII,Blakeslee2021_IRsb}. In recent years, the $H_0$ inference has raised tension between early-time probes, such as measurements from the cosmic microwave background and baryonic oscillations, and late-time probes using distance-ladder techniques \citep{Verde2019}. The time-delay technique, which consists of inferring the `time-delay distance' from the modelling of strongly lensed quasars for which time-delays have been measured \citep{Refsdal1964} is among the late-time probes, and confers the advantage of providing a distance estimate not anchored on a local distance ladder. However, the accuracy of this technique is currently based on an \textit{a priori} knowledge of early-type galaxies and of their mass density profiles \citep[e.g.][]{RXJ1131, TDLMC}. 

The main lensing galaxy mass distribution is often assumed to follow an elliptical power-law density profile or to be the sum of a baryonic component and dark matter elliptical density component. The role of the assumption of a radial density profile in the time-delay distance inference has been extensively discussed in the literature \citep{SS2013, SS2014, Xu2016, Kochanek2020, TDCOSMOIV,Kochanek2021}. 
Less attention has been given to quantifying the impact of an angular change of the density profile on the distance inference. \cite{suyu2010_B1608} quantified the impact of angular structures on $H_0$ for the time-delay lens system B1608+656 by allowing for a pixelized perturbation of an elliptical density profile. These authors showed that, for that system, perturbations of $<$ 2\% of the lens potential were required, yielding a change in $H_0$ of less than 1\%. This result motivated the use of mostly elliptical mass distribution in subsequent time-delay studies by the H0LICOW  and STRIDES collaborations. Nevertheless, a more generic investigation of the role of angular structure on $H_0$ is still missing. In the present work, we focus on specific angular structures: Fourier-like/multipole perturbations to the elliptical shape of the density profile. The study of early-type galaxies shows that their isophotes are well described by ellipses with Fourier deviations, also called multipoles, of orders three and four \citep[e.g.][]{Rest2001,Hao2006, Pasquali2006, Krajnovic2013,Mitsuda2017}. While the third-order components are generally of small amplitude, those of order four can be substantial, giving rise to clear discy or boxy shapes of the light profile. To our knowledge, the imprint left by multipoles present in the lensing galaxy on extended lensed images has not been discussed in detail in the literature, nor has its impact on $H_0$.

Due to lensing cross-section, massive early-type galaxies constitute the vast majority of the lens population discovered so far. They are commonly well modeled with density profiles that are perfectly elliptical. Apart from nearby galaxies explicitly included in the model, the only source of angular perturbation generically considered is tidal perturbation ---also known as external shear--- caused for example by other galaxies along the line of sight towards the lens. Those perturbations of the gravitational potential $\psi$ are quadrupolar, of the form $\psi \propto \cos{(2 \chi)}$, where $\chi$ is the azimuthal angle \citep{Trotter2000,Saas-Fee_part2_2006,Chu2013}. Higher order multipoles, corresponding to $\psi \propto \cos{(m \chi)}$, are often ignored as they are expected to displace lensed image positions by $\Delta \theta < 0.003\arcsec$ \citep[Sect. 4 in][]{Saas-Fee_part2_2006}. This forecast however relies on a rather small multipole amplitude ($a_4/\theta_E \sim 0.01$, where $\theta_E$ is the Einstein radius of the considered lens), and on the assumption that the light provides a good proxy of the perturbation existing in the total mass. 

Only a few studies have included an investigation of the constraints offered by lensing observables on those multipoles. \cite{Trotter2000} used very long baseline interferometry (VLBI) observations of multiple structures in MGJ0414+0534 to constrain high-order multipoles ($m$=3, 4) in that system. \cite{Claeskens2006} also used the compact star formation regions observed in J1131-1231 to constrain $m$=3, 4 terms of the multipole expansion. In both cases, the physical meaning of those multipoles was undetermined as either their amplitude or direction could hardly fit any reasonable physical model. The role of multipoles in introducing scatter around the fundamental surface of quad lenses \citep{Gomer2018} was recently investigated, revealing that the multipoles may not be sufficient to explain the observed deviation of lensed quasar positions from this surface. The presence of high-order multipoles can also modify flux ratios between quasar images  \citep[e.g.][]{Moller2003,Winn2003,Keeton2003,Keeton2005}. Those perturbations to the flux ratios introduced by macro-structures in the density profile need to be accounted for when using flux ratio anomalies to constrain the halo mass function \citep[e.g.][]{Mao1998, Xu2015, Gilman2019, Hsueh2020,Gilman2021}. Observational constraints on the amplitude of the high-order multipoles associated to the total mass distribution would therefore be highly valuable, as they could be used  to quantify the impact of low-amplitude boxyness and discyness on flux ratio anomalies.

With the advent of deep, high-angular-resolution images of lensed systems, as achievable with the Hubble Space Telescope (HST) or with adaptive optics systems, it is now possible to detect extended images of distant lensed galaxies, even those hosting a quasar \citep[e.g.][]{SLACSI,SHARPI,H0LiCOWI}. The Einstein rings are known to provide numerous constraints on the mass model and the extended source shape \citep{Kochanek2001}. The only systems for which the shape of the Einstein ring has been used to constrain the amplitude of multipoles are SDSS J0924+0219, HE 0435-1223, B1938+666, and PG 1115+080 \citep{Yoo2005,Yoo2006}. For each galaxy, the deviation of the mass density from ellipsoidal shape because of fourth-order multipoles was found to be consistent with zero, and only upper limits on $a_4$ could be derived. 
Extended lensed images are now commonly reproduced by lensed models down to the noise, even when the density profile of the lensing galaxy is modelled as an elliptical power-law density profile  \citep{RXJ1131,Shajib_magnitudes}.

The primary aim of this paper is to assess the detectability of multipolar components, and our second goal is then to quantify the extent to which they modify the $H_0$ inference. By quantifying the detectability of multipoles in high-resolution lensed images, we intent to find out if multipole-component-induced deformations are effectively negligible in those systems or have been absorbed by the model. In the latter case, we quantify the deviation of the time-delay distance induced by the fitted elliptical model.
While this is the main motivation of this work, the presence of multipoles in the total mass profile is also important for our general understanding of galaxies and of the ability of interactions between baryons and dark matter to shape the galaxy density profile. It is also relevant for dark matter studies based on flux ratio anomalies \citep{Hsueh2017,Hsueh2020,Gilman2020BirrerNierenberg,Gilman2020DuBenson,Nierenberg2020,Gilman2021}: 
the ability to constrain the presence of (dark) multipole-like structures in the lens from the analysis of the lensed images would enable further tightening of the constraints on the dark matter properties inferred from flux ratio anomalies. 

In Sect. \ref{section_method}, we introduce our general methodology, the mathematical definitions of the main lens mass profiles used in our analysis, and explain mock-image creation and fitting procedure. In Sect. \ref{section_results}, we systematically test the detectability of multipoles depending on different parameters and discuss their detectability in real cases. In Sect. \ref{section_discu}, we then quantify the effect of the presence of  multipoles on $H_0$ inference at a single-galaxy level and at a population level and assess their potential influence in cosmological lensing analyses. We summarise and conclude in Sect. \ref{section_conclu}.

For all calculations presented in this paper involving non-angular quantities, we assume a flat $\Lambda$CDM cosmological model with $\Omega_{\rm m}=0.3$, $\Omega_\Lambda=0.7$, and $H_0=70$ $\text{km}\,\text{s}^{-1}\,\text{Mpc}^{-1}$.

\section{Methodology}
\label{section_method}

To quantify the detectability of multipoles from strongly lensed images, we designed a controlled set of simulated lensed systems containing multipoles and modelled them using state-of-the-art techniques. Specifically, we emulate mock HST-like data of a quasar and its host lensed by a singular isothermal ellipsoid (SIE) with fourth-order multipoles and shear. Those multipoles are the dominant non-zero multipole contributions that control boxyness or discyness of galaxies (see Fig. \ref{boxy_discy_isokappa}). We then fit the mock images with an SIE and shear (i.e. without fourth-order azimuthal structures).

With this setup, we then analyse the effect of different factors on the quality of the retrieved fit. A good fit means that the multipoles remain unnoticed in imaging data of a lens system. This is the situation where multipoles may introduce unnoted systematic errors into the analysis, especially into the $H_0$ inference. Conversely, a poor fit would be an indication of modelling inadequacy. We assume that any structure detected in the residuals\footnote{We show in Sect.~\ref{section_results} that multipoles have rather characteristic imprints on the arc images, such that their origin may be easy to identify.} would be identified as a need to include multipoles in the lens  mass model. 
However, we do not explore the ability of macro-models to effectively recover the input multipoles. Instead, we seek the limit at which visible residual patterns become present over the noise. This depends on the mass and source model assumptions and on the data quality. Specifically, we explore the role of the shear, the lens axis-ratio, the additional freedom granted by relaxing the slope constraint in the fit ---and the influence of the radial mass profile in that case---, the signal-to-noise ratio (S/N), the multipole strength, and the freedom allowed in the quasar host galaxy morphology during the fit. 

We finally look at the impact of unnoticed multipolar components on time-delay cosmography. Focusing on our previous mock systems in cases where multipoles remain undetected, we first explore the $H_0$ recovery for those specific systems. 
We then extend the study by constructing a sample of lenses that emulates a realistic population of lenses and look at the combined $H_0$ inference for different S/N levels.

\subsection{Lens mass profile}
\label{sect_lens_model}

To construct mock images, one needs to assess a lensing mass profile. As we are interested in the effect of azimuthal structures such as boxyness and discyness in the lensing galaxy, we first select a radially simple mass profile.
The density profile of lensing galaxies is generally well approximated by a mass distribution that follows a power-law density profile and some external shear \citep{Suyu2009,suyu2010_B1608,review_tdc, yildirim2020,HolicowXIII}. The mean slope of the profile has been found to be close to isothermal \citep{livre_jaune,koopmans2003, suyu2010_B1608}, such that the singular isothermal ellipsoid (SIE) provides a simple but realistic choice of density profile for the lens\footnote{In Sects. \ref{influ_slope}--\ref{influ_compo}, more complex mass profiles are considered. The mathematical details of those profiles are given in Appendix \ref{appendix_mass_profile}. }. In addition, we add a multipole perturbation to the ellipticity and include a shear term to account for the tidal perturbation of the potential, such as the one due to line-of-sight mass distribution. Below, we describe the lensing potential associated to each of these components. The deflection angle and Hessian matrix can be found analytically by differentiation of the potential.

Following \cite{KeetonKochNIE}, the potential of our SIE is given by the potential of a non-singular isothermal ellipsoid (NIE) with a virtually null core size. 
This translates into a potential ---for an SIE aligned with the x-axis--- at position $\boldsymbol{\theta}=(x,y)$:
$$\psi_{\text{SIE}}(\boldsymbol{\theta}) = x \, \alpha_x+y \, \alpha_y ,$$
with 
$$\alpha_x(\boldsymbol{\theta}) = \frac{b}{\sqrt{1-q^2}} \arctan \left(\frac{\sqrt{1-q^2}x}{\sqrt{q^2x^2+y^2}}\right),$$
$$\alpha_y(\boldsymbol{\theta}) = \frac{b}{\sqrt{1-q^2}} \text{arctanh} \left(\frac{\sqrt{1-q^2}y}{\sqrt{q^2x^2+y^2}}\right),$$
where $q$ is the axis ratio, and $b$ is the scale radius and is equal to $\theta_E \sqrt{q}$, where $\theta_E$ is the Einstein radius. The difference in potential between two SIEs with different ellipticities is a quadrupole moment.

\begin{figure}
    \centering
    \includegraphics[width=0.48\textwidth]{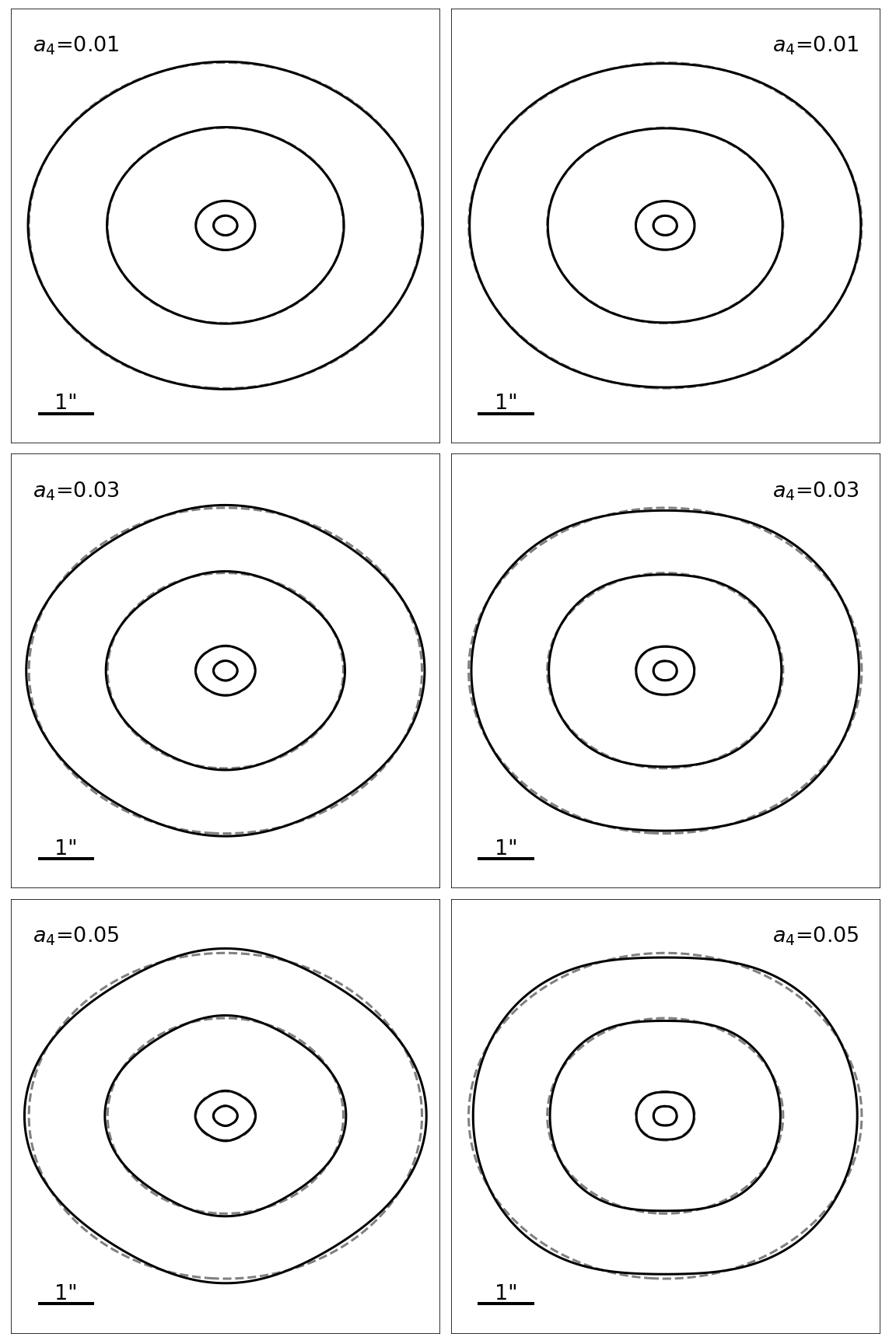}
    \caption{Isodensity contours for an SIE mass profile (grey dashed line) and an SIE + fourth-order multipole mass profile (black plain line). The multipole component has a strength $a_4=0.01$ (top), $a_4=0.03$ (middle), or $a_4=0.05$ (bottom) and is oriented along the SIE main axis to create a discy profile (left), or at 45$\degree$ to the SIE main axis to create a boxy profile (right). }
    \label{boxy_discy_isokappa}
\end{figure}

\begin{figure}
    \centering
    \includegraphics[width=0.49\textwidth]{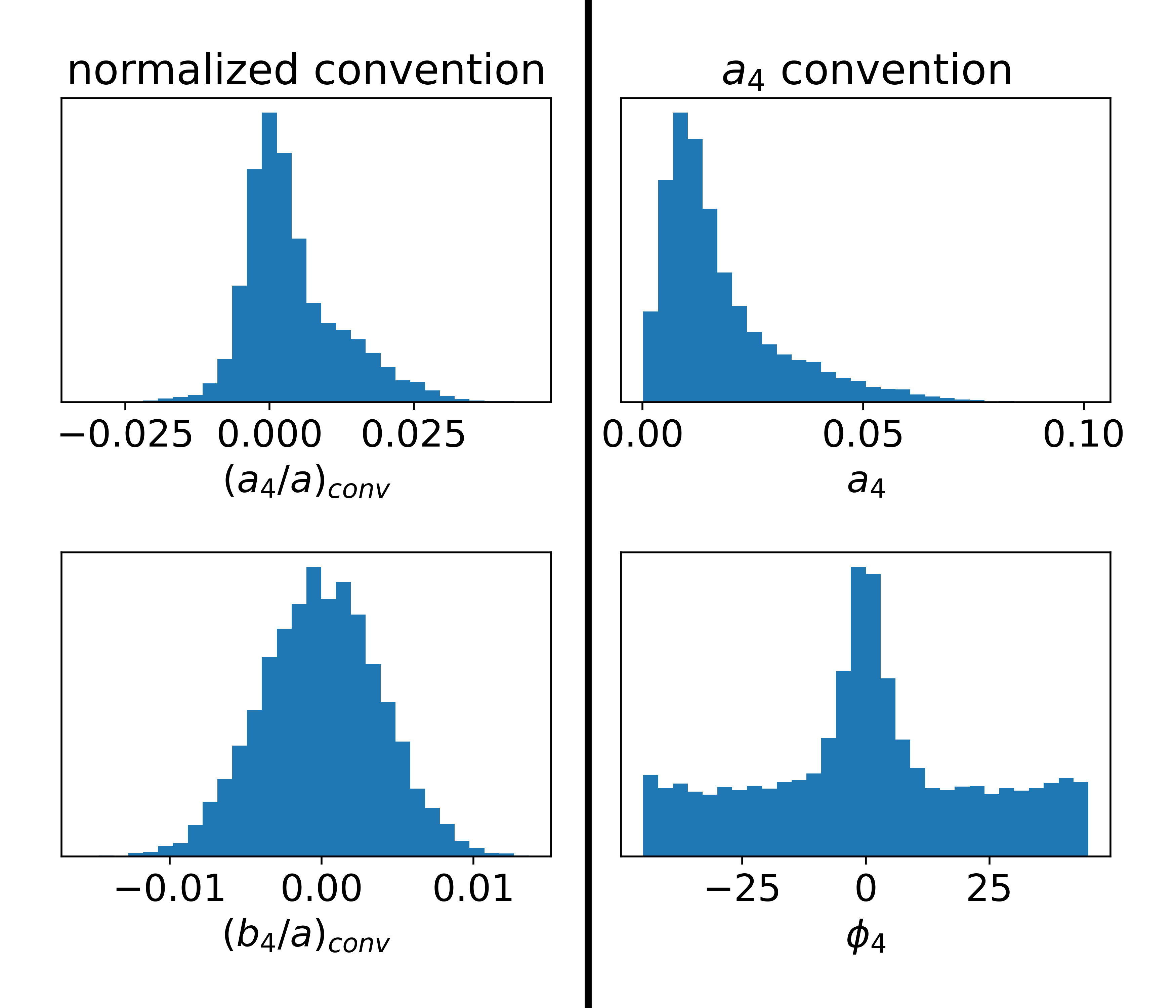}
    \caption{Distribution of fourth-order multipoles in  the light of local elliptical galaxies from \cite{Hao2006}. The two panels display the distribution for the normalised $(a_4/a)_{\text{conv}}$ and $(b_4/a)_{\text{conv}}$ convention (left) and our $a_4$ convention used in this paper (right), for $\theta_E=2\arcsec$ and $q=0.8$ (see Equations \ref{eq_a4}--\ref{eq_phi4}).}
    \label{a4_conv}
\end{figure}

The multipole has several definitions in the literature. We chose to use the same convention as \cite{Xu2015} and \cite{Keeton2003}. The lensing potential at a position $\boldsymbol{\theta}=(x,y)=(\theta \cos{\phi},\theta \sin{\phi})$ is given by:
$$\psi_{\text{multipole}}(\boldsymbol{\theta})= \theta \frac{1}{1-m^2} a_m \cos(m(\phi-\phi_m)),$$
where $m$ is the order of the multipole, $a_m$ is the multipole strength, and $\phi_m$ is the main axis orientation angle. The fourth-order multipole  is an octupolar moment of the potential. It introduces a `discy' shape when $\phi_4=0$ and a `boxy' shape when $\phi_4=45\degree$ (see Fig. \ref{boxy_discy_isokappa}). In the sample studied by \cite{Hao2006}, 70\% of elliptical galaxies display a multipole $a_4 < 0.02$ (see Fig. \ref{a4_conv}). We note that other authors \citep[e.g.][]{Bender1988,Bender1989,Hao2006,Penoyre2017,Frigo2019} prefer the normalised $a_4$ and $b_4$ convention, using $(a_4/a)_{\text{conv}}$ and $(b_4/a)_{\text{conv}}$ where $a$ is the length of the semi-major axis of the reference ellipse, $(a_4/a)_{\text{conv}}$ represents the cosine deformations of that ellipse, and $(b_4/a)_{\text{conv}}$ represents the sine deformations. 
This normalised convention is more appropriate when dealing with specific isophotes at different radii, while our $a_4$ convention is particularly useful for the creation of a general lens density profile that can be defined independently of the main lens shape. Conversion between the two conventions is explained in Appendix B2 of \cite{Xu2015} and is summarised here :
\begin{eqnarray}
    a_4 &=& \sqrt{(a_4/a)_{\text{conv}}^2 + (b_4/a)_{\text{conv}}^2} \times \frac{\theta_E (\arcsec)}{\sqrt{q}} \label{eq_a4} \\
    \phi_4 &=& \frac{1}{4} \times \arctan \left(\frac{(b_4/a)_{\text{conv}}}{(a_4/a)_{\text{conv}}} \right)
    \label{eq_phi4}
.\end{eqnarray}
As we only consider pure boxyness and discyness in the following analysis (i.e.  we always have $(b_4/a)_{\text{conv}} = 0$), the backward conversion is simply 
\begin{equation}
(a_4/a)_{\text{conv}} = \frac{ \pm a_4 \times \sqrt{q}}{\theta_E(\arcsec)}, \label{eq_a4conv}
\end{equation} which is positive in the discy case and negative in the boxy case.

The shear is defined by the lensing potential at position $\boldsymbol{\theta}=(x,y)$:
$$\psi_{\text{shear}}(\boldsymbol{\theta})= \frac{1}{2} (\gamma_1 x^2 + 2 \gamma_2 xy - \gamma_1 y^2),$$
where $\gamma_1$ and $\gamma_2$ are the components of the complex shear. The shear strength $\gamma_{\text{ext}} = \sqrt{\gamma_1^2 + \gamma^2_2}$ and its orientation $\phi_{\text{ext}} = \frac{1}{2}\arctan(\gamma_2 / \gamma_1)$. The shear is a quadrupole moment of the potential.

For our baseline model, we consider $\theta_E=2 \arcsec$ and $q=0.8$ for the SIE. We orient the SIE with an angle of $22\degree$ with respect to the x-axis. The orientation of the multipole $\phi_4$ is equal to either $0\degree$ or $45 \degree$ in a reference frame aligned with the main lens, and this will create either discy or boxy mass profiles. The strength of the multipole is $a_4=0.01$ by default. We fix the shear amplitude to 0.05 and its orientation to $30\degree$ with respect to the SIE main axis. The shear is therefore neither oriented on a multipole axis nor on the SIE main axis. No spurious artificial shear due to truncation of pixelated mass maps is
introduced \citep{VandeVyvere2020} because we have an analytical description of each component of the lens mass profile.

While one can argue that using an Einstein radius of 2$\arcsec$ is on the upper edge of the distribution of Einstein radii in observed lensing systems, it is particularly relevant to use such an $\theta_E$ in our study to highlight the effect of multipoles. As explained in Sect. \ref{other_influ}, for a fixed exposure time and source magnitude and shape, a smaller Einstein radius would yield a smaller signal-to-noise ratio in the ring and therefore reduce our ability to explore the lens properties that most impact the multipole detectability.

\subsection{Mock images creation}
\label{subsect_mock_crea}
The modelling of extended lensed images requires high-spatial-resolution data, which in general are provided by space-based or by ground-based AO systems \citep[e.g.][]{Lagattuta2010,Chen2016_1131,H0LiCOWI}. In this work, we chose to emulate mock\footnote{Notebook link: \faGithub \url{https://github.com/TDCOSMO/TD_data_public}} HST images observed with the WFC3 camera through the filter F160W. This choice is guided by the data used in many time-delay cosmography studies \citep{H0LiCOWI}, and in other studies using high-quality mock lens images \citep[e.g.][]{TDLMC,Park2021,WagnerCarena2021}. The near-infrared (NIR) band confers the advantage that, in general, it provides  the highest host brightness \citep[see e.g.][]{Ross2009}, and is \textit{a priori} the most favourable band with which to analyse extended lensed features. In addition, we also consider a transparent lens, which is equivalent to assuming that the lens light is perfectly subtracted and provides a negligible contribution to the photon noise in the ring. Due to the lack of lens light, this setup
slightly increases the S/N in the ring compared to a real NIR HST image. On the other hand, state-of-the-art lens modelling is based on multi-band HST data. This increases the effective signal in the ring. Moreover, the bluer bands have higher angular resolution and better sampling of the PSF due to smaller pixels, even if they are not able to fully balance the lack of host flux at those wavelengths. The relative advantage of the bluer bands is also that the lens light drops off more quickly, and therefore blends less with the ring. This last benefit is counterbalanced in our setup by the use of a transparent lens. Therefore, our single-band setup should, at least partially, compensate for the lack of bluer bands and give a reasonable indication of the detectability of multipole-induced deformation of the ring.

We used the same PSF as the one used in the Time Delay Lens Modeling Challenge \citep[TDLMC,][]{TDLMC} for the Rung 2 and 3, created from the drizzling of eight PSFs extracted from real HST images.
We use a pixel size of 0.08$\arcsec$, typical of drizzled NIR HST images. 
We include photon and read-out noise (RON). The zero-point in the AB system is 25.96 mag. Our baseline setup corresponds to an exposure of 5400s. The `sky' brightness, that is zodiacal light and Earth shine, is fixed at 22.3 mag/arcsec$^2$ and the RON  is set to 21 e$^-$. Those values are based on the WFC3 documentation \cite{WFC3} and concur with other works creating typical HST WFC3/F160W images \citep{TDLMC,Park2021,WagnerCarena2021}.

We consider a simple source light distribution constituted from an AGN on top of a circular host following a Sersic profile. The choice of a circular source is based on the fact that using a more complicated source with a principal axis may produce results that will be highly dependent on its orientation, while the goal of this analysis is to create mock systems where the azimuthal structure of the lens is the only source of potential bias. We set the contrast between the QSO and its host based on the work of \cite{Dunlop2003} and \cite{Jahnke2003}. This corresponds to a host that is brighter than the QSO by 0.5 mag. 
For our baseline model, the unlensed quasar apparent brightness is set to 21 mag and the unlensed host is set to 20.5 mag. This yields a mean magnitude of the lensed QSO of 18.5 mag. The host galaxy morphology is a circular Sersic with a half-light radius $R_{\rm sersic}=0.1 \arcsec$ and a Sersic index $n_{\rm sersic}=3$. The source is placed at a redshift $z_{s}=2$. 

\begin{figure}
    \centering
    \includegraphics[width=0.45\textwidth]{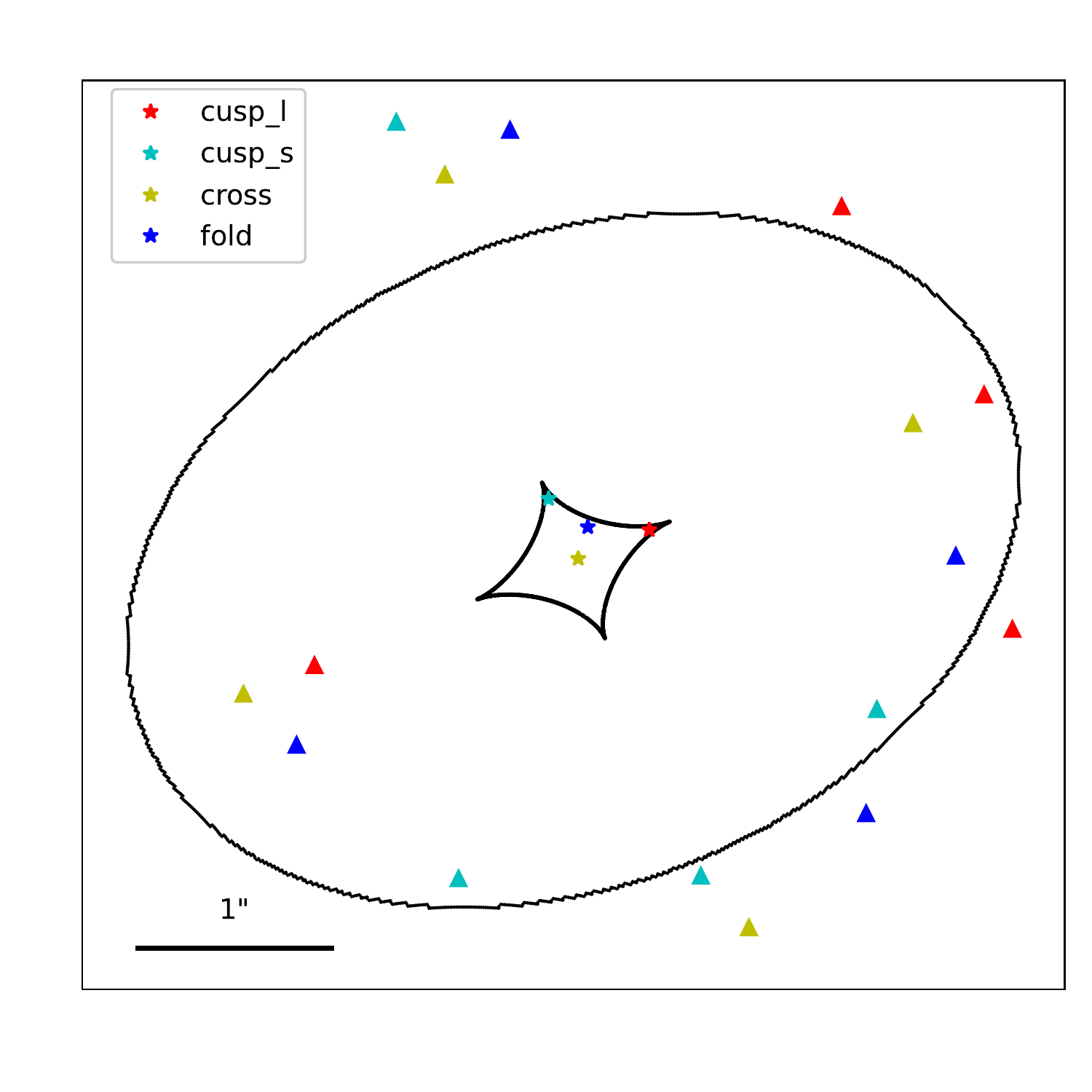}
    \caption{Illustration of the four lensing configurations used in this work. The inner caustic (bold) and associated critical curve are represented in black. The four different source positions are shown with star symbols and the corresponding image positions are drawn with triangles. Each configuration is displayed in a single colour (see legend)}
    \label{config_ex}
\end{figure}

The lensing galaxy density profiles are described in the previous section. The lens redshift is $z_l=0.271$. For each mock lens density profile, we only consider quad configurations. We focus on four-lensed-images configurations for two reasons: first, cosmographic studies prioritise such systems as they provide multiple delays. Second, dark matter investigations based on flux ratio anomalies currently only rely  on samples of quads. More specifically, we consider four quad configurations of lensed images by positioning the source at various locations with respect to the inner caustic. The four sources and associated image positions are illustrated in Fig. \ref{config_ex}. The source is either close to the long axis cusp (\textit{cusp\_l}), close to the short axis cusp (\textit{cusp\_s}), close to a \textit{fold} caustic, or centred (\textit{cross} image configuration). Specifically, we locate the source at 80\% of the distance between the centre and the cusp and fold caustics. The source for the cross-configuration is at a 5\% distance from the centre in the direction of the long axis cusp. The \texttt{lenstronomy} software\footnote{\faGithub ~ \url{https://github.com/sibirrer/lenstronomy}} is used to compute the mock images \citep{lenstro2018}.

The time-delays are calculated within \texttt{lenstronomy}. We fix the time-delays to the `ground truth' from the macro model but we consider a 2\% uncertainty on the delay  with a lower boundary of 1 day when performing the lens modelling. 

\subsection{Modelling setup}
\label{sect_modeling_setup}
Once the mock image is created, we model it with a standard modelling technique constrained by the pixels inside a mask which omits all pixels that are below twice the background noise level. We use \texttt{lenstronomy} and its particle swarm optimisation \citep[PSO,][]{PSO1995,PSO1998} to find the model parameters that maximise the likelihood. The likelihood contains terms associated to the image residuals, to the time-delays, and to the position of the four ray-traced back point-source positions in the source plane \citep[see][]{lenstro2018}. This last likelihood term allows efficient sampling by relaxing the constraint that all point sources should ray-trace back to the exact same position whilst still demanding a sufficiently accurate matching of the lens equation. The model considered is identical to the input mass profile except that no multipoles are included: the lens model is an SIE + shear, the source model is a circular Sersic, and the four point-sources are modelled in the image plane with their positions in the source plane matching the source centre. After the PSO has converged, a Markov Chain Monte Carlo (MCMC) using \texttt{emcee} \citep{MCMC_emcee2010,emcee2013} is performed to sample the posterior probability distribution.

\section{Multipole detectability}
\label{section_results}

\begin{figure*}
    \centering
    \fcolorbox{red}{red}{
    \includegraphics[width=0.98\textwidth]{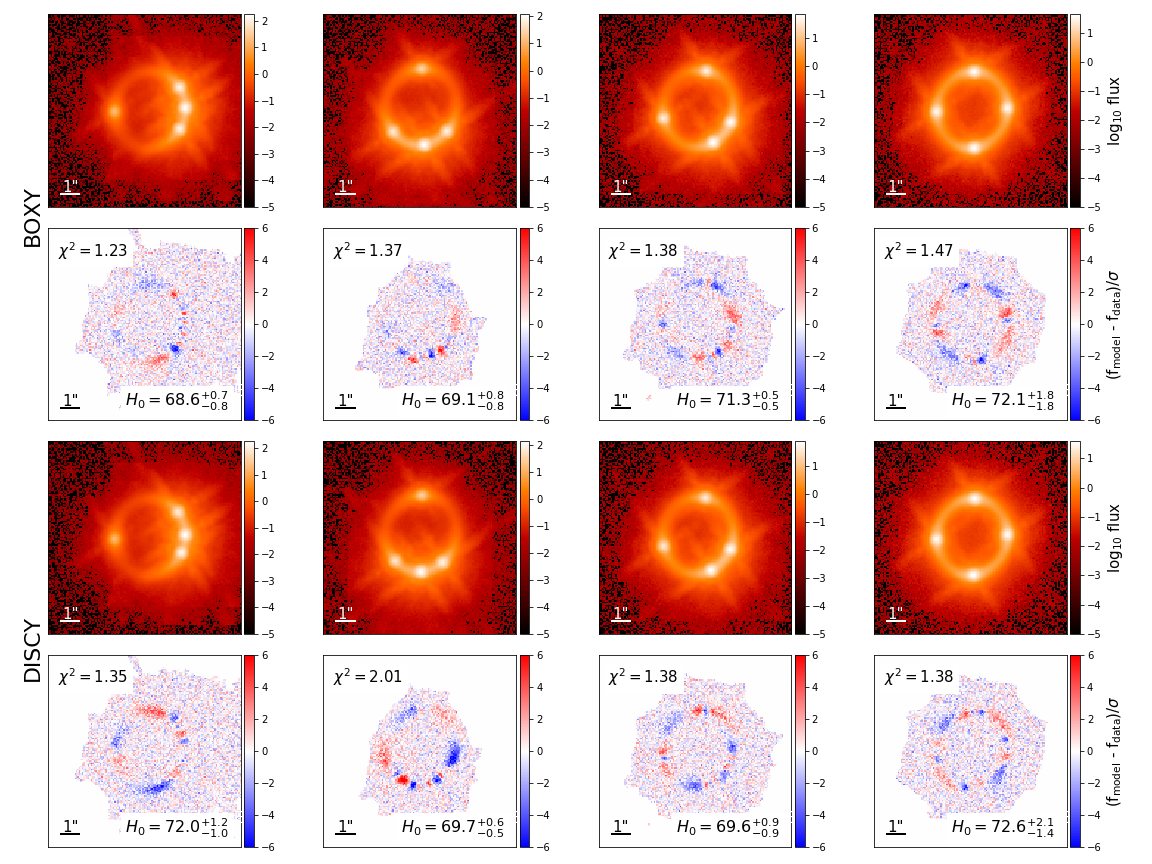}}
    \caption{Results of the base case mock images created and fitted as described in Sect. \ref{section_method}. Those systems are also displayed in red in Table \ref{big_result_table}. Top 2 rows: Mock images made from one boxy galaxy for four configurations and residuals associated to the fit for each image with $\chi^2$ and $H_0$ inference. Bottom 2 rows: Same as in the top rows but for the discy lensing galaxy.}
    \label{base_case}
\end{figure*}

We first study the base case described in the previous section. 
We set the multipole to $a_4=0.01$ and $\phi_4=0\degree$ with respect to the SIE main axis for discy cases and $\phi_4=45\degree$ with respect to the SIE main axis for boxy cases; see top panel of Fig. \ref{boxy_discy_isokappa}. This multipole level of boxyness and discyness is typical of what is observed in the light profile of elliptical galaxies (Fig.~\ref{a4_conv}). 
The resulting fits are shown in Fig. \ref{base_case}. The residuals and $\chi^2$ displayed are for the best model, that is, the model with the highest likelihood. The $H_0$ results are the median value of the MCMC sampling with the error corresponding to the 0.16 and 0.84 quantile. Regardless of the image configuration and multipole angle, we see an alternation of positive and negative residuals beyond the 3$\sigma$ level when running azimuthally along the ring. Those correlated structures are sufficiently large and numerous to yield a reduced $\chi^2$ much larger than 1. 
The impact on $H_0$ inference is discussed in Sect. \ref{section_discu}.

\begin{figure*}
    \centering
   \fcolorbox{ForestGreen}{ForestGreen}{
   \includegraphics[width=0.98\textwidth]{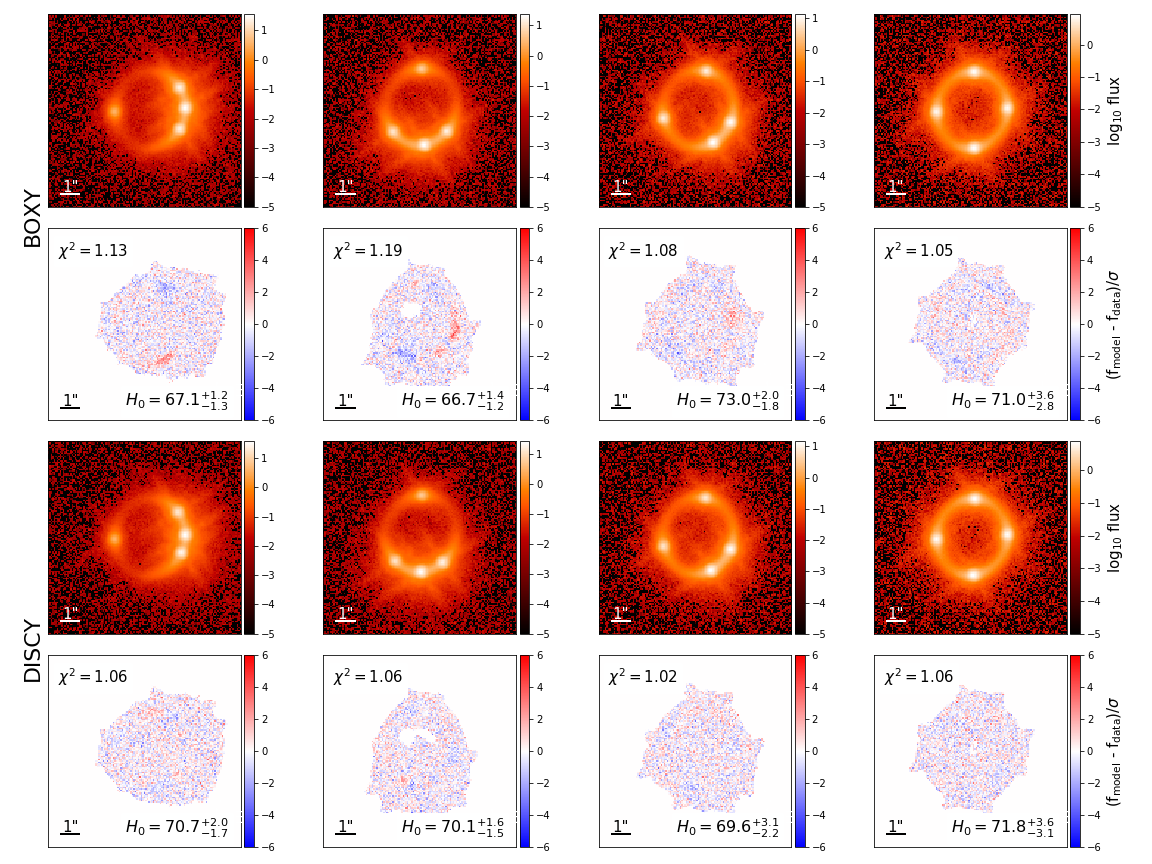}}
    \caption{Same as Fig. \ref{base_case}, but with an exposure time of 3000s instead of 5400s and the unlensed quasar and extended source fainter by 1.75 mag. Those results are also displayed in green in Table \ref{big_result_table}. We note that the mask, which is defined by a threshold of twice the background noise, may sometimes have holes inside the Einstein ring due to the lack of lensed signal at those locations.}
    \label{med_S/N}
\end{figure*}

\begin{figure}
    \centering
    \fcolorbox{Orchid}{Orchid}{
    \includegraphics[width=0.45\textwidth]
    {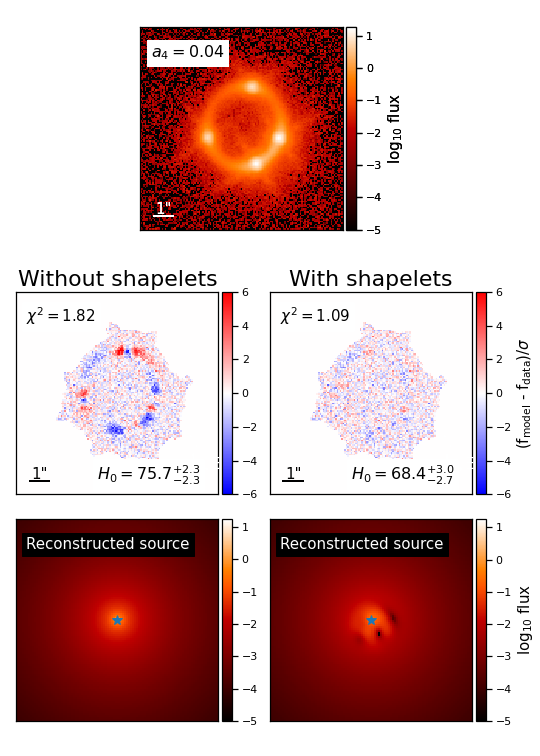}}
    \caption{Illustration of the influence of complexity in the modelled source. Top: Mock image created for 3000s exposure time and a bright source. The lens has the same characteristics as for the base case (see Sect. \ref{sect_lens_model}), but is discy with $a_4=0.04$ in a fold configuration. Bottom: Fit of the above image with (right) and without (left) adding shapelet reconstruction for the source during the fitting process. These results are also displayed in purple in Table \ref{big_result_table}.}
    \label{res_shapelets_shap}
\end{figure}

Our baseline setup shows that the imprint of multipoles on lensed images would be ubiquitous. We now systematically investigate how deviation from the baseline assumption impacts those results. We quantify the influence of the following parameters:
\begin{itemize}
    \item Shear: we investigate the extreme case where the shear amplitude of the mock system is equal to zero. 
    \item Lensing galaxy ellipticity: we consider three values representative of existing lensing galaxy samples, $q = (0.7,0.8,0.9)$.
    \item Assumption on the fitted lens mass profile: we allow the power-law slope to deviate from the isothermal case in the modelling while keeping the input slope isothermal.
    \item Assumption on the fitted lens mass profile when the mock lensing galaxy is not a SIE: we extend the previous point by using a mock lensing galaxy with varying profile slope with radius, following a baryon + dark matter prescription, and allow the fitted power-law slope to deviate from the isothermal case.
    \item S/N: we test three different S/Ns by varying the exposure time and/or the source and quasar brightness.
    \item Multipole strength: we vary $a_4$ between 0.01 and 0.05. We note that comparison of those values to multipole amplitudes found in the literature may require the use of \eqref{eq_a4conv}.
    \item Assumption of the source morphology: we include shapelets in the source reconstruction during the modelling. 
    
\end{itemize}

The following sections discuss the results obtained for each setup change. Table \ref{big_result_table} summarises the main results of those tests, providing for each case the value of the reduced $\chi^2$ for the best model and inferred value of $H_0$ from the MCMC sampling (median, 0.16 quantile, and 0.84 quantile).
Two subsamples of mock systems appear multiple times in the table: our baseline case (red) and a lower S/N (3000\,s exposure time with a bright source) case with $a_4 = 0.01$ (green). Our base case is the ideal case for which  multipoles would leave ubiquitous imprints in the lensed images (see Fig. \ref{base_case}); it is used as the starting point to compare the influence of different parameters such as the shear, the galaxy ellipticity, and the S/N. Because imprints from multipoles are ubiquitous with the fiducial case, we have built a lower S/N system to study other factors (see Fig. \ref{med_S/N}). This lower S/N setting is afterwards used for studying the influence of the multipole strength and the addition of source structure (i.e. shapelets) in the modelling. A specific pair of tests illustrating how the source can absorb perturbations introduced by multipoles is shown in Fig. \ref{res_shapelets_shap}. The corresponding cases are outlined in purple in Table~\ref{big_result_table}.

\begin{table*}
    \centering
    
\begin{tabular}{cc|c|c|c|c|c|c|c|c|}

                         & & \multicolumn{2}{c|}{cusp\_l}                                                                      & \multicolumn{2}{c|}{cusp\_s}                                                                      & \multicolumn{2}{c|}{fold}                                                                         & \multicolumn{2}{c|}{cross}                                                                        \\ \cline{3-10} 
\multirow{-2}{*}{}        && Boxy                                            & Discy                                           & Boxy                                            & Discy                                           & Boxy                                            & Discy                                           & Boxy                                            & Discy                                           \\ \hline \hline

&$\chi^2$&\color{red}$ 1.23$&\color{red}$ 1.35$&\color{red}$ 1.37$&\color{red}$ 2.01$&\color{red}$ 1.38$&\color{red}$ 1.38$&\color{red}$ 1.47$&\color{red}$ 1.38$\\  
\multirow{-2}{*}{with shear}&$H_0$&\color{red}$68.6^{+0.7}_{-0.8}$&\color{red}$72.0^{+1.2}_{-1.0}$&\color{red}$69.1^{+0.8}_{-0.8}$&\color{red}$69.7^{+0.6}_{-0.5}$&\color{red}$71.3^{+0.5}_{-0.5}$&\color{red}$69.6^{+0.9}_{-0.9}$&\color{red}$72.1^{+1.8}_{-1.8}$&\color{red}$72.6^{+2.1}_{-1.4}$ \sr \\ \hline 
&$\chi^2$&$ 1.22$&$ 1.03$&$ 1.35$&$ 1.03$&$ 1.34$&$ 1.31$&$ 1.31$&$ 1.38$\\  
\multirow{-2}{*}{without shear}&$H_0$&$69.9^{+0.8}_{-0.9}$&$70.2^{+0.8}_{-0.6}$&$71.8^{+1.1}_{-1.0}$&$70.4^{+0.6}_{-0.7}$&$69.8^{+1.5}_{-1.4}$&$73.3^{+1.0}_{-1.1}$&$66.3^{+1.1}_{-1.1}$&$75.3^{+1.6}_{-1.6}$ \sr \\ \hline \hline

&$\chi^2$&$ 1.09$&$ 1.04$&$ 1.32$&$ 1.31$&$ 1.15$&$ 1.16$&$ 1.11$&$ 1.09$\\  
\multirow{-2}{*}{$q=0.7$}&$H_0$&$71.2^{+1.1}_{-1.2}$&$71.1^{+1.1}_{-1.1}$&$69.7^{+1.0}_{-0.8}$&$71.0^{+0.7}_{-0.6}$&$71.7^{+0.8}_{-0.8}$&$70.8^{+0.7}_{-0.9}$&$70.6^{+2.1}_{-2.5}$&$73.7^{+1.5}_{-1.5}$ \sr \\ \hline 
&$\chi^2$&\color{red}$ 1.23$&\color{red}$1.35$&\color{red}$ 1.37$&\color{red}$ 2.01$&\color{red}$ 1.38$&\color{red}$ 1.38$&\color{red}$ 1.47$&\color{red}$ 1.38$\\  
\multirow{-2}{*}{$q=0.8$}&$H_0$&\color{red}$68.6^{+0.7}_{-0.8}$&\color{red}$72.0^{+1.2}_{-1.0}$&\color{red}$69.1^{+0.8}_{-0.8}$&\color{red}$69.7^{+0.6}_{-0.5}$&\color{red}$71.3^{+0.5}_{-0.5}$&\color{red}$69.6^{+0.9}_{-0.9}$&\color{red}$72.1^{+1.8}_{-1.8}$&\color{red}$72.6^{+2.1}_{-1.4}$ \sr \\ \hline 
&$\chi^2$&$ 1.17$&$ 1.42$&$ 1.62$&$ 1.28$&$ 1.82$&$ 2.00$&$ 2.65$&$ 2.69$\\  
\multirow{-2}{*}{$q=0.9$}&$H_0$&$70.4^{+0.8}_{-0.7}$&$72.6^{+2.2}_{-2.8}$&$74.1^{+0.8}_{-0.8}$&$69.4^{+0.7}_{-0.8}$&$73.2^{+1.1}_{-0.9}$&$70.4^{+1.0}_{-0.7}$&$75.9^{+0.9}_{-1.2}$&$71.0^{+1.1}_{-1.0}$ \sr \\ \hline \hline

&$\chi^2$&\color{red}$ 1.23$&\color{red}$ 1.35$&\color{red}$ 1.37$&\color{red}$ 2.01$&\color{red}$ 1.38$&\color{red}$ 1.38$&\color{red}$ 1.47$&\color{red}$ 1.38$\\  
\multirow{-2}{*}{slope fixed}&$H_0$&\color{red}$68.6^{+0.7}_{-0.8}$&\color{red}$72.0^{+1.2}_{-1.0}$&\color{red}$69.1^{+0.8}_{-0.8}$&\color{red}$69.7^{+0.6}_{-0.5}$&\color{red}$71.3^{+0.5}_{-0.5}$&\color{red}$69.6^{+0.9}_{-0.9}$&\color{red}$72.1^{+1.8}_{-1.8}$&\color{red}$72.6^{+2.1}_{-1.4}$ \sr \\ \hline 
&$\chi^2$&$ 1.20$&$ 1.23$&$ 1.33$&$ 1.63$&$ 1.38$&$ 1.35$&$1.41$&$ 1.28$\\  
\multirow{-2}{*}{\textbf{slope free}}&$H_0$&$63.3^{+1.6}_{-1.7}$&$80.0^{+2.3}_{-2.4}$&$65.1^{+1.1}_{-1.1}$&$79.3^{+1.4}_{-1.3}$&$74.6^{+2.2}_{-2.5}$&$75.7^{+2.8}_{-2.3}$&$79.7^{+4.8}_{-3.9}$&$84.4^{+4.5}_{-4.2}$ \sr \\ \hline \hline

&$\chi^2$&\color{red}$ 1.23$&\color{red}$ 1.35$&\color{red}$ 1.37$&\color{red}$ 2.01$&\color{red}$ 1.38$&\color{red}$ 1.38$&\color{red}$ 1.47$&\color{red}$ 1.38$\\  
\multirow{-2}{*}{5400s, very bright source}&$H_0$&\color{red}$68.6^{+0.7}_{-0.8}$&\color{red}$72.0^{+1.2}_{-1.0}$&\color{red}$69.1^{+0.8}_{-0.8}$&\color{red}$69.7^{+0.6}_{-0.5}$&\color{red}$71.3^{+0.5}_{-0.5}$&\color{red}$69.6^{+0.9}_{-0.9}$&\color{red}$72.1^{+1.8}_{-1.8}$&\color{red}$72.6^{+2.1}_{-1.4}$ \sr \\ \hline 
&$\chi^2$&\color{ForestGreen}$ 1.13$&\color{ForestGreen}$ 1.06$&\color{ForestGreen}$ 1.19$&\color{ForestGreen}$ 1.06$&\color{ForestGreen}$ 1.08$&\color{ForestGreen}$ 1.01$&\color{ForestGreen}$ 1.05$&\color{ForestGreen}$ 1.06$\\  
\multirow{-2}{*}{3000s, bright source}&$H_0$&\color{ForestGreen}$67.1^{+1.2}_{-1.3}$&\color{ForestGreen}$70.7^{+2.0}_{-1.7}$&\color{ForestGreen}$66.7^{+1.4}_{-1.2}$&\color{ForestGreen}$70.1^{+1.6}_{-1.5}$&\color{ForestGreen}$73.0^{+2.0}_{-1.8}$&\color{ForestGreen}$69.6^{+3.1}_{-2.2}$&\color{ForestGreen}$71.0^{+3.6}_{-2.8}$&\color{ForestGreen}$71.8^{+3.6}_{-3.1}$ \sr \\ \hline 
&$\chi^2$&$ 1.04$&$ 1.04$&$ 1.06$&$ 1.03$&$ 1.03$&$ 1.00$&$ 1.08$&$1.03$\\  
\multirow{-2}{*}{3000s, faint source}&$H_0$&$64.6^{+1.7}_{-1.3}$&$70.3^{+2.2}_{-1.8}$&$66.1^{+1.4}_{-1.3}$&$70.3^{+1.5}_{-1.5}$&$71.2^{+2.4}_{-2.4}$&$70.2^{+2.3}_{-2.4}$&$72.1^{+4.0}_{-3.8}$&$73.7^{+3.1}_{-3.6}$ \sr \\ \hline 
 \hline

&$\chi^2$&\color{ForestGreen}$ 1.13$&\color{ForestGreen}$ 1.06$&\color{ForestGreen}$ 1.19$&\color{ForestGreen}$ 1.06$&\color{ForestGreen}$ 1.08$&\color{ForestGreen}$ 1.01$&\color{ForestGreen}$ 1.05$&\color{ForestGreen}$ 1.06$\\  
\multirow{-2}{*}{3000s bright, $a_4=0.01$}&$H_0$&\color{ForestGreen}$67.1^{+1.2}_{-1.3}$&\color{ForestGreen}$70.7^{+2.0}_{-1.7}$&\color{ForestGreen}$66.7^{+1.4}_{-1.2}$&\color{ForestGreen}$70.1^{+1.6}_{-1.5}$&\color{ForestGreen}$73.0^{+2.0}_{-1.8}$&\color{ForestGreen}$69.6^{+3.1}_{-2.2}$&\color{ForestGreen}$71.0^{+3.6}_{-2.8}$&\color{ForestGreen}$71.8^{+3.6}_{-3.1}$ \sr \\ \hline 
&$\chi^2$&$ 1.36$&$ 1.15$&$ 1.83$&$ 1.18$&$ 1.32$&$1.16$&$1.31$&$ 1.17$\\  
\multirow{-2}{*}{3000s bright, $a_4=0.02$}&$H_0$&$59.6^{+1.9}_{-1.4}$&$70.1^{+1.7}_{-1.9}$&$64.7^{+1.3}_{-1.6}$&$71.7^{+1.3}_{-1.3}$&$74.6^{+1.9}_{-1.7}$&$9.9^{+2.7}_{-2.2}$&$73.8^{+3.2}_{-3.3}$&$77.8^{+5.8}_{-4.9}$ \sr \\ \hline 
&$\chi^2$&$1.64$&$ 1.27$&$ 2.87$&$ 1.29$&$ 1.64$&$ 1.45$&$ 1.77$&$ 1.29$\\  
\multirow{-2}{*}{3000s bright, $a_4=0.03$}&$H_0$&$55.7^{+0.9}_{-0.8}$&$71.9^{+2.3}_{-2.1}$&$69.5^{+1.6}_{-1.3}$&$73.3^{+1.4}_{-1.1}$&$77.3^{+2.0}_{-1.5}$&$71.6^{+2.2}_{-2.5}$&$78.9^{+5.4}_{-4.4}$&$76.7^{+4.9}_{-4.4}$ \sr \\ \hline 
&$\chi^2$&$ 1.87$&$ 1.34$&$ 4.36$&$ 1.66$&$ 1.97$&$\color{Orchid} 1.82$&$ 2.32$&$ 1.57$\\  
\multirow{-2}{*}{3000s bright, $a_4=0.04$}&$H_0$&$49.1^{+1.1}_{-1.0}$&$69.2^{+2.1}_{-1.5}$&$73.2^{+1.5}_{-1.1}$&$75.3^{+1.7}_{-1.3}$&$79.4^{+2.4}_{-2.5}$&\color{Orchid}$75.7^{+2.3}_{-2.3}$&$84.3^{+4.4}_{-3.9}$&$79.3^{+4.2}_{-3.1}$ \sr \\ \hline 
&$\chi^2$&$ 3.39$&$ 1.51$&$ 6.02$&$ 1.84$&$ 2.50$&$2.45$&$ 3.17$&$ 1.89$\\  
\multirow{-2}{*}{3000s bright, $a_4=0.05$}&$H_0$&$47.7^{+1.4}_{-1.3}$&$71.3^{+2.0}_{-1.9}$&$81.4^{+1.6}_{-2.0}$&$75.5^{+1.6}_{-1.5}$&$83.7^{+2.8}_{-2.7}$&$77.2^{+2.9}_{-2.4}$&$90.1^{+4.2}_{-4.4}$&$82.7^{+5.0}_{-4.1}$ \sr \\ \hline \hline

&$\chi^2$&$ 1.03$&$ 1.02$&$ 1.07$&$ 1.00$&$ 1.02$&$0.98$&$ 1.01$&$ 1.04$\\  
\multirow{-2}{*}{3000s bright, $a_4$=0.01,  \textbf{shapelet}}&$H_0$&$66.2^{+1.8}_{-2.0}$&$71.1^{+2.5}_{-2.1}$&$62.8^{+1.3}_{-1.5}$&$71.0^{+1.6}_{-1.7}$&$72.9^{+3.3}_{-2.8}$&$69.7^{+2.9}_{-3.1}$&$68.5^{+3.4}_{-3.2}$&$71.6^{+7.4}_{-5.0}$ \sr\\ \hline 
&$\chi^2$&$1.07$&$ 1.07$&$ 1.18$&$ 1.10$&$ 1.11$&$ 1.04$&$ 1.14$&$ 1.10$\\  
\multirow{-2}{*}{3000s bright, $a_4$=0.02, \textbf{shapelet}}&$H_0$&$62.3^{+2.0}_{-1.8}$&$71.6^{+2.0}_{-1.9}$&$58.2^{+1.3}_{-1.3}$&$71.9^{+1.5}_{-1.8}$&$72.1^{+2.6}_{-2.3}$&$69.1^{+3.0}_{-2.6}$&$66.0^{+3.3}_{-2.6}$&$77.2^{+5.7}_{-6.1}$ \sr \\ \hline 
&$\chi^2$&$ 1.13$&$ 1.10$&$ 1.22$&$ 1.16$&$ 1.21$&$ 1.18$&$ 1.31$&$ 1.16$\\  
\multirow{-2}{*}{3000s bright, $a_4$=0.03, \textbf{shapelet}}&$H_0$&$59.1^{+1.1}_{-1.0}$&$73.1^{+2.4}_{-1.9}$&$63.9^{+1.4}_{-1.6}$&$75.3^{+2.0}_{-1.4}$&$72.3^{+2.6}_{-2.4}$&$68.9^{+3.3}_{-2.6}$&$41.3^{+3.1}_{-3.3}$&$72.8^{+4.4}_{-4.0}$ \sr \\ \hline 
&$\chi^2$&$ 1.21$&$ 1.13$&$ 1.50$&$ 1.13$&$ 1.17$&\color{Orchid} $1.09$&$ 1.62$&$ 1.14$\\  
\multirow{-2}{*}{3000s bright, $a_4$=0.04, \textbf{shapelet}}&$H_0$&$54.7^{+0.8}_{-0.7}$&$67.4^{+2.3}_{-2.1}$&$61.3^{+1.6}_{-1.6}$&$77.0^{+1.9}_{-2.1}$&$69.6^{+2.1}_{-2.6}$&\color{Orchid}$68.4^{+3.0}_{-2.7}$&$73.2^{+2.4}_{-2.1}$&$77.2^{+6.2}_{-5.2}$ \sr \\ \hline 
&$\chi^2$&$ 1.47$&$1.19$&$ 1.63$&$ 1.44$&$ 1.26$&$ 1.12$&$ 2.25$&$ 1.47$\\  
\multirow{-2}{*}{3000s bright, $a_4$=0.05, \textbf{shapelet}}&$H_0$&$42.2^{+1.1}_{-1.2}$&$69.4^{+2.0}_{-1.9}$&$44.2^{+1.4}_{-1.2}$&$75.8^{+2.1}_{-2.1}$&$69.1^{+2.9}_{-2.4}$&$67.3^{+3.3}_{-2.4}$&$62.7^{+2.5}_{-2.3}$&$75.4^{+5.1}_{-4.3}$ \sr \\ \hline \hline

\end{tabular}
\caption{Summary table of $\chi^2$ and $H_0$ results for the cases tested in Sect. \ref{section_results}. }
\tablefoot{The base case has $\gamma_{\text{ext}}=0.05$, $q=0.8$, 5400\,s exposure time, a very bright source ($\langle m_{\rm{QSO}} \rangle$=18.5\,mag), and $a_4=0.01$ and is fitted without shapelets in the source. This setup appears at different places in the table; the corresponding rows are coloured in red and the results are also shown in Fig.~\ref{base_case}. The differences between the base case and the others are indicated in the first column, the rows list the different modifications applied to the base case. The specifications in \textbf{bold} are changes made with respect to the model, while the other changes are specific to the mock-image creation. The 3000\,s exposure time and bright source ($\langle m_{\rm{QSO}} \rangle$=20.25\,mag) setup with $a_4$=0.01 is displayed at two rows in green, the related residuals are shown in Fig. \ref{med_S/N}. The same setup with $a_4=0.04$, with and without shapelet reconstruction is displayed in purple. The associated residuals are shown in Fig. \ref{res_shapelets_shap}. The uncertainties on $H_0$ are calculated using the 0.16 and 0.84 quantiles and the main value is the median of the MCMC $H_0$ sampling.}
\label{big_result_table}

\end{table*}

\subsection{Influence of the shear}

We compare the modelling of our base case, which is characterised by $\gamma_{\text{ext}}=0.05$ and an orientation of $30\degree$ with respect to the SIE major axis, with systems generated without shear. We find that the absence of shear does not significantly change the goodness of the fit, apart from in two specific cases.

As shown in Table \ref{big_result_table}, the general trend is the inability of the shear to absorb deformations introduced by multipoles. This is expected as a fourth-order multipole  (i.e. $m=4$) is an octupolar moment while the shear is a quadrupolar one. However, two mock systems stand out: the discy ones with cusp configurations. For those two cases, we find that good residuals can be achieved. This may be explained by the specific symmetry: the multipole, the SIE, and the source position are aligned, thus creating images in a relatively narrow region with conserved axial symmetry (for the potential, magnification, and deflection). It is thus easier to mimic the effect of input multipole with an SIE$+$shear in this small symmetric region. In addition,  for those two systems, we observe that the fit of an SIE on an SIE$+$multipole is not equivalent to fitting an SIE$+$shear on SIE$+$shear$+$multipole. The former is not presented in the table but is a good fit, while the latter is a poor fit. This result may appear surprising at first sight as the same additional component (the shear) is added to the mock system and to the model. However, while the lensing potential, deflection, and convergence of the different components sum up linearly, the magnification is not linearly changed, thus explaining the difference between sheared and shearless cases.

\subsection{Influence of the lens ellipticity}
\label{sec_q}

Early-type lensing galaxies generally have axis ratios in the range $q \in $ [0.65-0.95] \citep[e.g.][]{Sluse2012,Rusu2016,Shajib_magnitudes}. We therefore tried three different axis ratios for the lensing galaxy: a more elliptical galaxy with $q=0.7$, the base case with $q=0.8$, and a rounder galaxy with $q=0.9$. We find that the rounder the galaxy, the poorer the fit (see Table \ref{big_result_table}). This means that the detectability of multipoles is increased when the galaxy is rounder. 

The main cause of this effect may be the following: as the ellipticity increases (i.e. $q$ decreases), the size of the inner caustic increases, such that the magnification of a source located at the same relative position with respect to the caustic gets smaller. For a fixed exposure time, this yields a lower effective S/N in the ring, reducing our ability to detect multipole imprints in the lensed host images.

Apart from this general trend, a careful look at Table \ref{big_result_table} shows that for a small number of cases the fit slightly improved with the increase in $q$. This may be due to the specific realisation of the noise, and the slight dependence of the results on the position of the source with respect to the caustics. We note that to test the impact of $q$, we compared systems holding the value of $a_4$ to 0.01. According to Eq. \ref{eq_a4conv}, this implies a different value of $a_4/a$ for the different $q$. Keeping $a_4/a$ constant instead of $a_4$ constant does not change the increase in $\chi^2$ with the axis ratio $q$.

\subsection{Influence of a relaxed slope constraint}
\label{influ_slope}

Above, we model the lens with an isothermal mass model, which is our fiducial lensing galaxy profile. However, a power law with a free slope is often used in the modelling of real lensing systems. We therefore followed the same fitting procedure using a power-law elliptical mass density profile \citep[PEMD;][]{Barkana1998SPEMD} with an additional step in the fitting that allows the slope to be free to vary. A PEMD is characterised by a 3D density profile $\rho(r) \propto r^{-\gamma'}$. This model is equivalent to an SIE for a power-law index $\gamma'$=2. See Appendix \ref{appendix_pl} for more details.

As can be seen in Table \ref{big_result_table}, a slightly lower $\chi^2$ is obtained when relaxing the slope. However, changing the slope of the modelled profile is not sufficient to absorb the residual patterns due to the multipole. There is therefore no evidence for a degeneracy between the modelled mass density slope of the lens and the input fourth-order multipoles.

The impact of input angular structures on a fitted monopole was recently discussed by \cite{Kochanek2021}. He explains how angular structures impact the Einstein ring and, if fit with too simple a model, can either produce large $\chi^2$ or drag the radially fitted model to an incorrect slope. Here we have a true lens mass profile that displays boxyness or discyness and fit it with an elliptical power law and shear. Because the multipoles only weakly change the mass within the Einstein radius, and octupole perturbations of the potential cannot be mimicked by a combination of quadrupolar components (i.e. ellipticity and shear), the freedom allowed to the fitted slope of the power law is not sufficient to absorb the multipole effects.

\subsection{Influence of a relaxed slope constraint when the mock lens follows a composite mass profile}
\label{influ_compo}

While we observe in the previous section that relaxing the constraint on the fitted slope does not significantly improve the fit, one may wonder whether or not this is due to the fact that the same family of models, i.e. power law, is used for both the mock image creation and the fit, thus not allowing any possible degeneracy linked to the slope. To test this hypothesis, we created mock lensing galaxies with a composite mass profile comprised of a baryonic Chameleon \citep{Dutton2011} profile and a Navarro-Frenk-White \citep[NFW][]{NavarroFrenkWhite1996} dark matter component. The Einstein radius of these new lensing galaxies is chosen to be 2$\arcsec$, as used for our fiducial mock systems. We parameterized our mass profiles such that the lensing galaxy is similar to those observed in massive ellipical lensing galaxies, with a dark matter fraction of 0.32 within the Einstein radius, with an axis ratio $q =0.8$ constant with radius, and a slope close to isothermal at the Einstein radius. The complete set of characteristics of our mock profile is given in Appendix \ref{appendix_cham}--\ref{appendix_nfw}.

This experiment being mainly an expansion of Sect. \ref{influ_slope}, we display the results of this test separately in Table \ref{table_composite_results}. The first row of Table \ref{table_composite_results} contains the results when no multipole is present in the composite lens mass; this is used as a fiducial case to ensure the reference $\chi^2$ is close to one \footnote{We note that the recovered value of $H_0$ in the absence of multipole is not centred on 70\,$\text{km}\,\text{s}^{-1}\,\text{Mpc}^{-1}$. This is mostly due to the mass sheet degeneracy \citep[MSD,][]{SS2013}. Without kinematics information and modelled light of the lensing galaxy, such degeneracy cannot be broken \citep{TDCOSMOIV,TDCOSMOV,TDCOSMOVI}.}. As already observed, when the input mass profile follows a SIE, the freedom allowed in the fitted slope only slightly improves the $\chi^2$. On the other hand, we also observe that the typical $\chi^2$ obtained with the power-law model is similar when the truth is a composite (Table \ref{table_composite_results}, rows 2-3) or a SIE (Table \ref{big_result_table}, rows 6-7). 

The ability of the power-law model to fit the composite mock system when $a_4$=0 demonstrates that the two models are degenerate. The addition of multipoles to the mock lenses leaves the radial profile and Einstein radius unchanged, such that the same region of the radial profile is probed by the host galaxy images. The similarity of the $\chi^2$ obtained with the composite and power-law mock mass profiles is consistent with that interpretation. 
On the other hand, the similarity between the $\chi^2$ for both the composite and SIE mock systems (with and without multipoles) informs us that the role of the radial mass profile is less important for $\chi^2$ than effects such as modelled source shape freedom (Sect. \ref{influ_shap}) or other input characteristics such as the S/N (Sect. \ref{influ_snr}).

We note that for more complex azimuthal structures, the choice of the radial profile may not be ignored. For example, if we consider a varying ellipticity with radius but the multipole kept constant, we do not exclude the possibility of greater interplay between the radial profile and the azimuthal structures: the imprint of the multipoles on the lensed ring gets stronger for a rounder galaxy, as seen in Sect. \ref{sec_q}. As a consequence, a mock galaxy displaying varying slope and ellipticity with radius could potentially behave differently from what we have seen here under multipolar perturbations. However, this level of complexity is beyond the scope of this paper.
\begin{table*}
    \centering
    \begin{tabular}{cc|c|c|c|c|c|c|c|c|}

                         & & \multicolumn{2}{c|}{cusp\_l}                                                                      & \multicolumn{2}{c|}{cusp\_s}                                                                      & \multicolumn{2}{c|}{fold}                                                                         & \multicolumn{2}{c|}{cross}                                                                        \\ \hline

&$\chi^2$& \multicolumn{2}{c|}{$1.00$}& \multicolumn{2}{c|}{$1.02$}& \multicolumn{2}{c|}{$1.02$}& \multicolumn{2}{c|}{$1.04$}\\

\multirow{-2}{*}{IN: composite, no multipole} &$H_0$ &\multicolumn{2}{c|}{$61.8^{+2.5}_{-1.8}$}&\multicolumn{2}{c|}{$64.8^{+1.9}_{-1.6}$}&\multicolumn{2}{c|}{$65.1^{+2.6}_{-2.1}$}&\multicolumn{2}{c|}{$65.4^{+5.6}_{-4.6}$}\\ 
\multirow{-2}{*}{MOD: \bf{free slope}}& slope & \multicolumn{2}{c|}{$1.94$} & \multicolumn{2}{c|}{$1.96$}& \multicolumn{2}{c|}{$1.97$}& \multicolumn{2}{c|}{$1.96$} \\
\hline

\multirow{-2}{*}{}        && Boxy                                            & Discy                                           & Boxy                                            & Discy                                           & Boxy                                            & Discy                                           & Boxy                                            & Discy                                           \\ \hline 

&$\chi^2$&$ 1.46$&$ 1.24$&$ 1.61$&$ 1.43$&$ 1.45$&$ 1.28$&$ 1.42$&$ 1.29$\\  
\multirow{-2}{*}{IN: composite, MOD: \bf{fixed slope}}  &$H_0$ &$62.0^{+2.1}_{-2.2}$&$64.2^{+1.9}_{-2.0}$&$63.6^{+1.4}_{-1.4}$&$65.5^{+1.3}_{-1.4}$&$65.0^{+2.6}_{-2.8}$&$64.8^{+2.6}_{-2.8}$&$62.5^{+4.8}_{-4.6}$&$68.1^{+5.8}_{-5.4}$\\ \hline 
&$\chi^2$&$1.44$&$1.18$&$ 1.52$&$ 1.35$&$ 1.43$&$ 1.25$&$ 1.38$&$ 1.22$\\  
\multirow{-2}{*}{IN: composite, MOD: \bf{free slope}} &$H_0$ &$49.4^{+1.7}_{-2.2}$&$83.5^{+2.5}_{-2.5}$&$50.1^{+1.2}_{-1.3}$&$85.6^{+1.8}_{-2.0}$&$60.0^{+2.1}_{-2.3}$&$73.9^{+3.0}_{-2.9}$&$53.8^{+4.8}_{-4.3}$&$84.0^{+6.6}_{-5.2}$\\ \hline

    \end{tabular}
    \caption{Results of the experiments performed in Sect. \ref{influ_compo}. }
    \tablefoot{The mock systems follow the base case with $\gamma_{\text{ext}}=0.05$, $q=0.8$, 5400\,s exposure time, and a very bright source ($\langle m_{\rm{QSO}} \rangle$=18.5\,mag), but differ by the main lens mass profile which is a composite mass profile made of a baryonic component and a dark matter component, instead of being a SIE. The case represented in the first row is multipole free while the two bottom rows display a multipole $a_4=0.01$, either in discy or boxy orientation. The mock images are fitted with a power-law model (hence the \textbf{bold} characters to differentiate it from the base SIE modelling, following the conventions of Table \ref{big_result_table}) with either a free slope or a fixed slope. In the latter case, the fixed value is the value retrieved in the no multipole case for each configuration.}
    \label{table_composite_results}
\end{table*}

\subsection{Influence of signal-to-noise ratio}
\label{influ_snr}

Summarising the S/N of images of a lensed system with a single number is a difficult task, as the point source and the extended images need to be accounted for. We therefore opted for a more qualitative criterion based on the exposure length and the brightness of the source. Our base case has an exposure time of 5400s and a very bright source which has an intrinsic brightness of 20.5 mag for the host and 21 mag for the quasar. This translates into integrated lensed magnitudes around 17 (17) mag for the four images of the host (quasar). Considering each image separately, the typical mean lensed image magnitude is 18.5 mag. 
We then reduce the exposure time to 3000\,s to match the exposure obtained typically with one orbit of HST. For that exposure time, we consider two intrinsically fainter sources: a first source that has a mean observed magnitude of typically 20.25 mag and another one of 21 mag. Those systems with lower S/N correspond to unlensed quasar brightnesses of respectively 22.75 mag and  23.5 mag, and are referred to as `bright' and `faint' source cases below. While this latter case corresponds to a particularly faint quasar image, the corresponding host brightness is sometimes observed in lensed systems \citep{Ding2021_constrast}.

From Figs. \ref{base_case}--\ref{med_S/N} and Table \ref{big_result_table}, we see that there is a clear decrease in the prevalence of the residuals as the data quality decreases. Indeed, the highest S/N fits (our base case, in red in the table) display ubiquitous patterns in the residuals and the $\chi^2$ is always above 1.2. For the 3000s exposure time and bright source case (green in the table), we predominantly see cases for which the multipole imprint is hidden in the noise, but there are a few systems with $\chi^2$ close to 1.2 with light imprints. This is no longer the case for the faint source cases, for which the $\chi^2$ is consistently lower than 1.1.

The exposure time and source brightness are not the only factors that control the S/N, but they provide a reasonable observation-based proxy of the data quality. Other factors, such as the Einstein radius, the contrast between quasar and host, and so on, also influence the S/N, as discussed in Sect. \ref{other_influ}.

\subsection{Influence of the multipole strength}

To characterise the dependence of the fit on the input multipole strength, we created mock mass profiles for four additional values of $a_4 \in [0.01, 0.05]$. Figure \ref{boxy_discy_isokappa} provides a visual representation of the induced deviation from a pure elliptical profile. As the residuals for $a_4=0.01$ for our baseline setup have already been detected, we chose to vary the strength of the multipole for our second setup corresponding to an exposure time of 3000s and a bright source (see Sect. \ref{influ_snr}). The latter case with $a_4=0.01$ is shown above in Fig. \ref{med_S/N} and the green line in Table \ref{big_result_table}. If we convert the multipole convention of \cite{Hao2006} (eq. \ref{eq_a4conv}) to our notation, we see that 70\% of the galaxies of their sample have $a_4<0.02$ and only 4\% of their sample have $a_4>0.05$ (see Fig. \ref{a4_conv}). The low multipole cases ($a_4 \leq 0.02$) are therefore the most likely to be representative of a real sample of galaxies, while the case $a_4 = 0.05$ should be seen as a rare extreme situation that could plausibly happen in a large sample (typically 100) of time-delay lenses. We may note that boxy and discy galaxies are not evenly distributed, with 64\% of the sample by \cite{Hao2006} being discy. However, the relative rate of boxyness or discyness is also found to depend on the galaxy mass or brightness \citep{Hao2006,Kormendy2009,Mitsuda2017}.

As one would expect, it is less easy to achieve a good fit as the multipole strength increases (see Table \ref{big_result_table}). More precisely, if we assume a threshold of $\chi^2=1.2$ to distinguish between poor and good residuals, multipoles get detected above $a_4$=0.03 for discy galaxies while $a_4=0.02$ is the detection limit for boxy galaxies. 

\subsection{Influence of complexity in the modelled source}
\label{influ_shap}
When modelling real systems, the extended source is hardly ever modelled as one simple light profile \citep[e.g.][]{RXJ1131,J1206}. Most of the time, the source is either pixelised with regularisation applied \citep{Brewer2008,Suyu2010_source,Nightingale2015,Galan2021} or analytical with a main simple profile plus a set of basis functions, such as shapelets, that are designed to capture morphological host complexity beyond axisymmetry \citep{birrer_shapelets,RXJ1131,lenstro2018,J1206,Shajib_magnitudes}. 
Above, we model the source with a circular Sersic profile because we know that this is its true input profile. However, more flexibility in the source should be allowed, as in state-of-the-art modelling of strongly lensed quasars. Due to our choice of modelling software, we chose to add shapelets to the source \citep[as e.g.][]{RXJ1131,J1206,Shajib_magnitudes}. We performed an extra step in the fitting procedure of all the systems presented in the previous sections, adding shapelets to the source with maximal polynomial order $n_{\text{max}}=5$. 

As shown in Table \ref{big_result_table}, the improvement when shapelets are added in the source is ubiquitous. When the fit is already good without shapelets, that is, typically cases with $a_4=0.01$ for the bright source, there is only a slight improvement of the $\chi^2$. When the fit without shapelets displays large residual features, that is, typically when $\chi^2>1.2$, the improvement introduced by shapelets is more significant, as illustrated in Fig. \ref{res_shapelets_shap}. The retrieved source is generally more structured than the input circular one but is not unphysical (e.g. the source surface brightness remains positive). 

The approximate degeneracy between the source morphology and the main lens potential is likely an expression of the source position transformation \citep[SPT;][]{SS2014,Unruh2017,Wagner2018,Wertz2018}. Additional complexity in the source can compensate for complexity in the lensing potential, which has further-reaching effects in general than the specific exploration of multipoles in this work. In practice, the complexity of the source is unknown, but it could be informed using a sophisticated prior that emulates the known morphologies of host galaxies. As an example, the dark spots in the source reconstructions for our fits may be mathematically acceptable (in the sense that they have non-negative brightness values), but the human eye can recognise them as being unusual and may suspect that they are instead absorbing the complexity of the potential. With no straightforward way to quantify and implement this human intuition (or certainty as to whether it can always be trusted), identification of whether or not the source reconstruction corresponds to the intended quantity is a difficult task.

With our tests, we find that shapelet reconstruction can absorb some of the multipole complexity. However, one could also wonder whether or not allowing even more freedom in the source could absorb any additional imprint of lens complexity. While answering this question is not the main purpose of this work, we concisely explored two different ways of allowing more complexity in the source: allowing ellipticity in the Sersic source, and allowing shapelets with a maximal polynomial order of eight instead of five (both maximal orders were investigated in \cite{J1206}). After testing a few `extreme cases' with $\chi^2>1.3$, we found that the ellipticity in the source cannot compensate the lensing potential complexity. The improvement of the fit thanks to this new degree of freedom is negligible. On the other hand, shapelets up to order $n_{\text{max}}=$8 yield a substantial improvement of the $\chi^2$ compared to the fits obtained with $n_{\text{max}}=5$. Indeed, for the last row of Table \ref{big_result_table}, five of the eight mock images can be fitted with $\chi^2 < 1.2$ with $n_{\text{max}}=8$, while only two of the eight good fits can be obtained with $n_{\text{max}}=5$. This improvement in the image residuals comes at a cost in the source plane, namely the appearance of clumpy structures in the source that may hopefully be identified as unrealistic (see Appendix \ref{appendix_shap}). In summary, the freedom in the source reconstruction helps to absorb some of the multipole complexity, but it seems that the associated distortion of the source may not systematically mimic structures that exist in real galaxies. Future work is needed to more thoroughly explore the constraining power and limitations of an arbitrary number of shapelets.

\subsection{Other influences}
\label{other_influ}

Apart from the factors explored in the previous sections, there are yet other ingredients that could influence the detectability of multipoles, such as the Einstein radius, the source intrinsic shape, and the lens light contamination. We did not systematically quantify their impact, but we explain  their influence qualitatively.

As the Einstein radius gets smaller, the different images of the same background source are brought closer together and can easily overlap, especially in cusp cases. The extended source images consist of fewer pixels and the quasar images, with the PSF spreading their light and covering the ring, dominate most of the flux. As the multipolar component is visible in residuals mainly thanks to the host flux, having less host pixels reduces the ability to detect imprints from multipole perturbations. This effect can even be stronger than a change of S/N as performed in Sect. \ref{influ_snr}. However, precisely quantifying the role of the Einstein radius remains difficult as the detectability of multipolar structures in the ring becomes much more dependent on the lensed-images configuration, source shape, PSF extension, and contrast for smaller Einstein radii.

Similarly, if the contrast increases between the extended host flux and the quasar flux (i.e. the host is fainter for a given quasar magnitude), the multipolar components are less detectable at fixed exposure time. In addition, the source radial profile also plays a role: as the host displays a shallower profile (decrease of $n_{\text{sersic}}$), the total brightness is the same but the host is more rapidly lost in the noise. This therefore decreases the total S/N in the host flux and the multipole is less detectable. Following the same lines, if a source is larger (i.e. increase of $R_{\text{sersic}}$) and keeps the same magnitude in the source plane, the host image spreads over more pixels with less flux in each, also decreasing the S/N in the ring. We did not explore the influence of the source azimuthal shape (e.g. an elliptical source). We expect the effect to be less straightforward than a change in $n_{\text{sersic}}$ or $R_{\text{sersic}}$. We anticipate a complex dependence of the multipole detectability on the orientation of the source  relative to the caustics.

In our analysis, we considered a transparent lens (i.e. no lens light). However, the presence of the lens light may have two different effects. First, the lens light and the source images can be blended, making it difficult to resolve the two components: part of the flux attributed to the lens might in fact come from the host images perturbed by multipoles in the lens mass. To avoid blending issues between source and lens light, it is common to use images of the same lens system in different spectral bands \citep{Shajib_magnitudes}. We did not simulate any lens light in our analysis and therefore did not test the efficiency of multiband imaging to mitigate the potential issue arising from blending either. Second, the lens light in itself could potentially help in detecting multipole presence as multipolar components in the mass are linked to those in the light. The detectability of multipole in the lens light profile is discussed in more detail in Sect. \ref{detectability_real}.
The above list may not be exhaustive, but encompasses the main sources of effects that could affect the results presented in this work. 
\subsection{Detectability in real lens systems}
\label{detectability_real}
We see in the previous sections that several factors influence the goodness of the fit of our mock images. For the highest quality data, the imprint of the multipole on the lensed images is ubiquitous. However, there are circumstances where the multipole imprint can remain hidden, such as if shear is absent and the main axes of both the ellipticity and multipole are aligned with the source position. Multipole perturbations are also more difficult to detect in galaxies with a higher ellipticity (i.e. small axis ratio). Leaving the mass profile slope to vary improves the residuals but only slightly, even when the true input mass profile slope varies with radius. The S/N on the other hand is crucial for detecting multipoles: with low S/N, the multipole contribution can be totally concealed in the noise. Of course, it also depends on the strength of the multipole contribution and on the freedom allowed in modelling the source, for example using a shapelet basis on top of an analytical model. Other factors such as the contrast, the source radial profile, or the Einstein radius also add more complexity to the problem. 

Overall, the different factors that can increase or decrease the detectability of multipoles combine together and create a unique lensing system each time. It is not possible to characterise each system with a single number that would reflect its robustness to multipole detectability because it depends on so many parameters. 
However, it is possible to compare a system to our three different S/N levels.
For our three S/N levels as defined by exposure time and source brightness, we essentially find three tiers of multipole detectability: 1. multipole will never go unnoticed in the residuals, 2. multipole can go unnoticed if it is at a low level or absorbed through freedom in the source structure, and 3. multipole will nearly always go unnoticed. However, exposure time alone is not a complete descriptor of these thresholds, because a lens with for example a lower exposure time but a brighter extended source relative to the quasar (or larger Einstein radius, as another example) will mimic a higher level of multipole detectability. Consequently, in order to accurately quantify the detectability of multipoles in a given system,  the system must be precisely emulated so as to classify it in one of the described tiers.  

A complementary path that can be followed to assess the role of multipoles in the model is to  search for evidence of Fourier modes in the luminosity profile of the lens. \cite{Mitsuda2017} and \cite{Pasquali2006}  demonstrated that multipolar perturbation of the light profiles can be measured in elliptical galaxies at redshifts similar to those of lensing galaxies. It is \textit{a priori} less clear whether or not such a measurement is feasible in lensing galaxies for which the  lens light is blended with the light from the lensed images. To assess the feasibility of such a measurement in lensed systems, we used $F814W$ HST and $K$-band AO images of the lensed quasar \object{QSO J1131-1231} \citep{sluse2003_1131, suyu2013_1131, Chen2016_1131}. While the large Einstein ring ($\theta_E \sim 1.8\arcsec$) and bright lensing galaxy may ease the detection of multipoles in the lens, the lensed images are particularly bright with features extending down to the very inner region of the lens \citep[see e.g. Fig. 4 of ][]{Claeskens2006}. This makes this system a good test case. We performed multipole measurements on the images subtracted from the best ring model published by \cite{suyu2013_1131} and \cite{Chen2016_1131} using two different techniques. First, we fitted ellipses with fourth-order Fourier components using the Python package \texttt{photutils}. The results obtained on HST and AO data are compatible with each other for radii $0.4 < \theta < 0.9 \arcsec$. They suggest a mean value of $a_4 < 0.02$. At radius $\theta < 0.4\arcsec$, there is an insufficient number of pixels to enable any sensible measurement. At radius $\theta > 0.9\arcsec$, the residuals from the PSF and/or lower S/N in the lens prevented the algorithm from converging. For AO data, measurements were possible up to $\theta=2\arcsec$, confirming $a_4 \leq 0.02$. We used a second method consisting in fitting the full lens intensity profile with a Sersic model using \texttt{galfit} \citep{Peng2010_galfit}, but including a fourth-order Fourier mode  to describe the azimuthal profile of the lens. While the fit of a single Sersic profile provides a poor description of the profile in the inner regions of the lens \citep[see e.g.][]{Claeskens2006}, the amplitude $a_4 = 0.01$ agrees for the two data sets. However, the same is not true for $\phi_4$ which is offset in the two data sets by $30\deg$. We may note that these values of $a_4$ do not appear unrealistic from visual inspection of the images, as $a_4 > 0.04$ are easily detectable by eye in good S/N data, but no deviation from a perfect ellipse is apparent. In summary, this test demonstrates that a measurement of $a_4$ in the light of lensing galaxies may be feasible in high-resolution images, but more work is required to assess the amplitude of systematic errors in such a measurement. 

 Finally, we note that we did not test the inclusion of multipoles in the lens mass model. This is an alternative path that could also be explored, but is beyond the scope of this paper.

\section{Impact on $H_0$ inference.}
\label{section_discu}
We show in the previous section that azimuthal perturbations of the potential, if present, could remain unnoticed in lensed images. Importantly, those perturbations can be absorbed in the lens and/or source parameters, which then differ from their true values. 
In this section, we quantify their impact on the value of $H_0$ inferred from those (biased) models. While the goodness of the fit was the main topic of the previous section, we discuss here those results in terms of the $H_0$ inference. 
\subsection{At a single-galaxy level}

\begin{figure}
    \centering
    \includegraphics[width=0.45\textwidth]{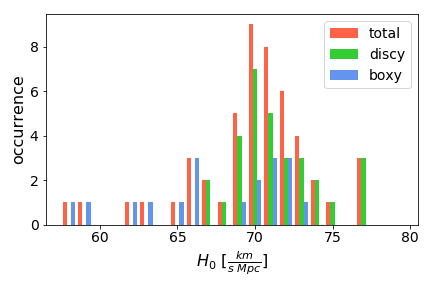}
    \caption{Median $H_0$ inference for mock systems from Table \ref{big_result_table} with $\chi^2 < 1.2$ (see text for more information). The histogram for boxy galxies is displayed in blue, that for discy ones is in green, and the sum of both is in red. The histograms are representative of the retrieved $H_0$ values obtained in our systematic tests that encompass multipole amplitudes of plausible lenses. However, this does not represent an inferred $H_0$ distribution from a realistic population of lenses (see Fig. \ref{pdf_final_h0}).}
    \label{h0_chi12}
\end{figure}

During our systematic tests shown in Table \ref{big_result_table}, the inferred $H_0$ values range typically between 60 and 80 $\text{km}\,\text{s}^{-1}\,\text{Mpc}^{-1}$ while the fiducial value is 70 $\text{km}\,\text{s}^{-1}\,\text{Mpc}^{-1}$. We disregard the results obtained for models yielding a poor $\chi^2$, as we consider that those models would not be included in a subsequent analysis. We also disregard the tests from Table \ref{table_composite_results} because the results are redundant with those of Sect. \ref{influ_slope} already present in Table \ref{big_result_table}. Figure \ref{h0_chi12} summarises the inferred $H_0$ values obtained for $\chi^2<1.2$. All the cases listed in Table \ref{big_result_table} fulfilling this criterion are included but we exclude duplicates. Specifically, we consider only once the coloured rows that appear multiple times in the table and only the shapelet reconstruction for the cases with 3000s exposure time and a bright source. In our sample, the discy and boxy galaxies do not behave in the same way: the discy galaxy models display values of $H_0$ of around 70 $\text{km}\,\text{s}^{-1}\,\text{Mpc}^{-1}$ and above; while the boxy ones show $H_0$ values with an extended tail below the fiducial $H_0$. If we consider only fits with $\chi^2<1.1$, the same trends are observed. As our sample is neither homogeneous nor highly representative of a  galaxy population, Fig. \ref{h0_chi12} is not the distribution of $H_0$ we would have derived if the value of $a_4$ of the lensing galaxies were randomly drawn from the distribution observed in a population of elliptical galaxies. However, each $H_0$ inference is representative of a possible elliptical galaxy.

We may now wish to quantify the impact of multipoles on $H_0$ when considering a population of lens systems. For this purpose, we first need to quantify the probability of having lensing galaxies deviating from a pure elliptical profile. For this purpose, we need first to identify if there is any evidence for a dependence on $a_4$ on the galaxy luminosity, mass, environment, and/or whether $a_4$ varies as function of the galacto-centric radius and of the redshift. We look at those various aspects in the following (Sects. \ref{sect_a4_elli_gal} and \ref{sect_population_level}).

\subsection{The $a_4$ distribution for elliptical galaxies}
\label{sect_a4_elli_gal}
\cite{Hao2006} studied the light of 847 galaxies at redshift $z<0.05$. They found that boxy galaxies tend to be bigger and brighter than discy ones and that discy galaxies are more abundant, with abundance ratios depending on the magnitude cutoff. The boxy galaxies are also more likely to be found in denser environments compared to discy ones. This corroborates the findings of \cite{Bender1988,Bender1989}  in several ways, except that the latter found equivalent abundances of boxy and discy galaxies. A search for a redshift evolution of $a_4$ was performed by \cite{Pasquali2006} and \cite{Mitsuda2017}. \cite{Pasquali2006} studied a small sample of 18 galaxies with $0.5<z<1.0$ and found more discy galaxies but suggested that this might be due to their limited sample. They corroborate previous findings that discy galaxies have higher ellipticities while boxy galaxies are more luminous, wider, and less elliptical in general. \cite{Pasquali2006} also found that the galaxies in their sample are similar to the ones at $z=0$ and only a possible trend in $a_3$ with redshift was observed. \cite{Mitsuda2017} extended this work by studying 130 galaxies at $z\sim1$ and 355 at $z\sim0$. These authors found, independently of the redshift, that galaxies showing $a_4$ contributions are very common and that discy shapes are favoured for galaxies with $log(M_*/M_{\odot})<11.5,$ while the boxy galaxies start to dominate at higher masses. \cite{Mitsuda2017} also looked at the $a_4$ radial profile of the disc-shaped NGC4697 in more detail: in the innermost region, the $a_4$ parameter is more uncertain and in some filters switches to a boxy profile. For the other parts of the galaxy, the $a_4$ is more or less constant with radius. The change of $a_4$ with galacto-centric radius has been studied more extensively by several authors \citep[e.g.][]{Rest2001,Kormendy2009,Krajnovic2013}. The first two authors studied the largest sample of elliptical galaxies. They find a great diversity in $a_4$ variation with radius. A profile can be boxy in the centre and discy in the outskirts, or be relatively constant or even display a monotonic increase or decrease. Discyness or boxyness can therefore exist at the regions encompassed by typical Einstein rings, that is mostly the region within 1 - 2 effective radii. In summary, at redshifts, masses, and ellipticities of galaxies considered for lensing, the presence of galaxies with nonzero boxyness or discyness is common. 

As the mass is not directly observable, we need to rely on numerical simulations to quantify the presence of such structures in the mass distribution. No authors have yet reported the presence of $a_4$ components in the total mass distribution, that is both stars and dark matter. However, the stellar mass dominates at radii $\lessapprox R_{Ein}$, and the studies focusing on multipolar components in the sellar mass are thus relevant for our purpose.  \cite{Penoyre2017} studied stellar mass distribution in the Illustris simulations \citep{Illustris} and found a few galaxies with clear boxyness and discyness but could not conclude on precise $a_4$ measurements due to too low spatial resolution. \cite{Frigo2019} presented `zoom' simulations and studied the boxyness and discyness of their sample more thoroughly. They did not find clear boxyness in their sample. They found systematic discyness if the galaxies are viewed edge-on. When seen at random orientations, the galaxies are still systematically discy but with lower $a_4$ values. The AGN feedback in their simulations tends to create less azimuthal structures at $z=0$ while the AGN influence is not predominant at $z=1$ where both AGN and no-AGN cases presented discyness, especially for more elliptical galaxies. 

In summary, we do not find any evidence for multipoles to be less prominent in galaxies characteristic of lensing galaxies, namely galaxies with high mass and z > 0.2, than in close-by galaxies. 
In conjunction, we see that for a single galaxy, realistic amplitudes of the multipoles can introduce a bias on $H_0$ (Fig. \ref{h0_chi12}). In the following section, we quantify the impact of multipoles on $H_0$ derived from a population of lensing galaxies.

\subsection{At the population level}
\label{sect_population_level}

In this section, we use our mock systems and a plausible distribution of $a_4$ to quantify their influence on the inferred distribution of $H_0$ from an ensemble of lens systems.
First, we repeat the $H_0$ inference for values of $a_4$ in the range $[0.00,0.05]$ \footnote{We note that systems with $a_4=0.00$ are only created once instead of twice for different $\phi_4$: they are valid for both discy and boxy statistics.}, a fixed axis ratio q=0.8, and the three different levels of S/N obtained by modifying the exposure time and source brightness (see Sect.~\ref{influ_snr}). This way, we expect to cover a representative range of S/N achievable in observations. As explained in Sect. \ref{detectability_real}, those three S/N levels are more broadly representative of three tiers of multipole detectability (almost always detectable; sometimes detectable; almost never detectable). 

\begin{figure*}
    \centering
    \includegraphics[width=0.48\textwidth]{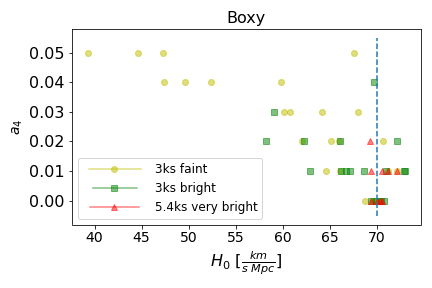}
    \includegraphics[width=0.48\textwidth]{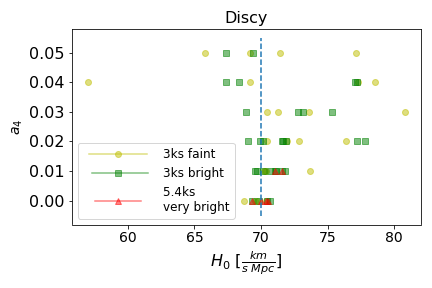}
    \caption{$H_0$ inference from mock systems created with different $a_4$ and different S/N (yellow circles : 3000s faint source ; green squares: 3000s bright source ; red triangles : 5400s very bright source). The boxy and discy  cases are in the left and right panels, respectively. The typical uncertainties representing the 0.16 and 0.84 quantiles are displayed in the legend. The fiducial $H_0$ is 70 $\text{km}\,\text{s}^{-1}\,\text{Mpc}^{-1}$ and is represented with the blue dashed line.}
    \label{a4vsH0_levels}
\end{figure*}

We then proceed to the fit of those mock systems with an SIE+shear model, as outlined in Sect. \ref{sect_modeling_setup}. When the fit is not good enough, that is when $\chi^2>1.2$, we add shapelets in the source light. Results are displayed in Fig. \ref{a4vsH0_levels} (considering only cases with $\chi^2<1.2$). 
As already seen in the previous section, poor fits are generally obtained for the highest S/N considered, such that only a few cases appear on the figure. The inferred value of $H_0$ tends to be underestimated when the galaxy is boxy and overestimated when discy. Moreover, the lower the S/N, the larger the spread in the $H_0$ inference. We note that the typical uncertainties reported in Fig. \ref{a4vsH0_levels} being the mean of the 68\% credible intervals, not only strongly depend on the S/N but also on the configurations: cross configurations yield uncertainties roughly twice as large compared to the other configurations. Thus, having more (less) good fits for crosses will lengthen (shorten) the final typical error bars.

\begin{figure*}
    \centering
    \includegraphics[width=0.48\textwidth]{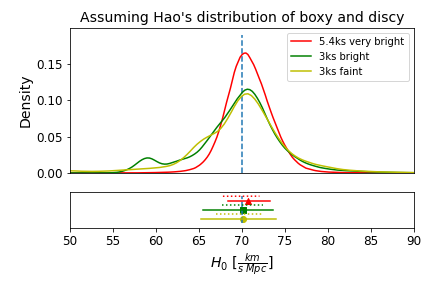}
    \includegraphics[width=0.48\textwidth]{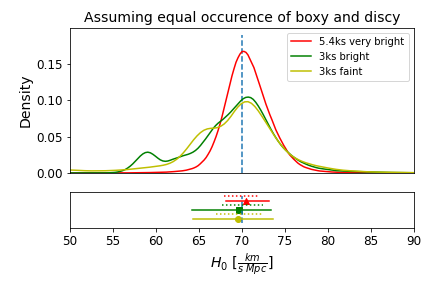}
    \caption{Underlying distribution of $H_0$ for our base case (5400s, very bright source) (red), and for 3000s bright (green) and faint (yellow) sources. 
    Each panel corresponds to two different fractions of boxy and discy galaxies (left: 36\% boxy, 64\% discy, as reported in \cite{Hao2006} ; right : 50\% boxy, 50\% discy). The corresponding median and 68\% credible interval are displayed below in solid lines. The dotted lines are the 68\% credible interval for the multipole-free distribution. It can be seen in this figure, as well as in Table \ref{table_final_pdf_h0}, that the presence of multipoles becomes a significant source of uncertainty for lower S/N.}
    \label{pdf_final_h0}
\end{figure*}

To forecast the distribution of values of $H_0$ for a population of discy and boxy galaxies similar to known early-type galaxies, we combine the results from Fig. \ref{a4vsH0_levels}, weighting each data point based on the distribution of $a_4$ derived by \cite{Hao2006} (Fig. \ref{a4_conv}). The distribution of $a_4$ is discretized into six bins: the bin with $a_4\leq0.005$ contains 10\% of the sample, the $0.005<a_4\leq0.015$ bin contains 46\%, the $0.015<a_4\leq0.025$ bin takes 21\%, the one with $0.025<a_4\leq0.035$ contains 10\%, the $0.035<a_4\leq0.045$ bin contains 6\%, and the $0.045<a_4\leq0.055$ bin contains 3\%. We neglect the rare systems with $a_4$>0.055. For each bin, the probability density function (PDF) is the average of the different PDFs associated to each point on the graph showing $a_4$ versus $H_0$  (Fig. \ref{a4vsH0_levels}) for a given type (boxy or discy) and a given $a_4$ value. Those PDFs are then weighted with the probability of encountering the $a_4$ value and by the fraction of discy and boxy galaxies assumed in the sample. The final PDF is thus 
\begin{equation}
\label{eq:PDF}
    \text{PDF}(H_0)=\sum_{i=1}^{6} \text{PDF}(i) \times \text{w}(i),
\end{equation}
where $i$ is the index corresponding to the $a_4$ bin, PDF$(i)$ is the associated PDF for each $a_4$ bin and type (boxy or discy), and w$(i)$ is the weight associated to each $a_4$ bin multiplied by the relative fraction of boxy and discy galaxies. In \cite{Hao2006}, discy galaxies represent 64\% of the sample while 36\% of it are more boxy. However, \cite{Mitsuda2017} show that the fraction of boxy/discy galaxies depends on the stellar mass of those galaxies. When $\log(M_*/M_\odot) > 11.5$, boxy and discy galaxies are equally  observed. The results for the two relative distributions of boxy and discy galaxies are reported in solid lines in Fig. \ref{pdf_final_h0} and the median and 68\% credible intervals are written in Table \ref{table_final_pdf_h0} in the $H_0$ columns. Those distributions correspond to the underlying $H_0$ distributions from which samples are drawn. In other words, from a sample of $n$ lensing galaxies, one is expected to infer $n$ independent measurements of $H_0$ that would follow the observed distribution.

\begin{table*}
\centering
\begin{tabular}{l|c|c|c|c|c|c|}
           & \multicolumn{3}{c|}{Hao's ditribution  }                                                                                             & \multicolumn{3}{c|}{50/50 distribution}                                                                                               \\ 
\cline{2-7}
           & \multicolumn{1}{c|}{$H_0$}  & \multicolumn{1}{c|}{$\sigma_{\text{multipole}}$} & \multicolumn{1}{c|}{$\sigma_{\text{no multipole}}$} & \multicolumn{1}{c|}{$H_0$}  & \multicolumn{1}{c|}{$\sigma_{\text{multipole}}$} & \multicolumn{1}{c|}{$\sigma_{\text{no multipole}}$}  \\

           & \multicolumn{1}{c|}{[$\text{km}\,\text{s}^{-1}\,\text{Mpc}^{-1}$]}  & \multicolumn{1}{c|}{[\%$H_0$]} & \multicolumn{1}{c|}{[\%$H_0$]} & \multicolumn{1}{c|}{[$\text{km}\,\text{s}^{-1}\,\text{Mpc}^{-1}$]}  & \multicolumn{1}{c|}{[\%$H_0$]} & \multicolumn{1}{c|}{[\%$H_0$]}  \\ 
\hline

5400s very bright   & $70.7^{+2.6}_{-2.4}$ & 1.7\%                                           & 3.0\%                                              & $70.5^{+2.7}_{-2.3}$ & 2.0\%                                           & 2.9\%                                            \Tstrut{}  \srb \\
3000s bright & $70.1^{+3.5}_{-4.6}$ & 4.7\%                                           & 3.4\%                                              & $69.7^{+3.7}_{-5.5}$ & 5.6\%                                           & 3.4\%                                               \srb \\
3000s faint    & $70.2^{+3.8}_{-4.9}$ & 4.9\%                                           & 3.8\%                                              & $69.6^{+4.0}_{-5.2}$ & 5.4\%                                           & 3.9\%       \srb                                       
\end{tabular}
\caption{$H_0$ distributions from Fig. \ref{pdf_final_h0}} 
\tablefoot{$H_0$ distributions for the different S/Ns and the two different fractions of boxy and discy galaxies. left: 36\% boxy, 64\% discy, following \cite{Hao2006} ; right: 50\% boxy, 50\% discy. For each distribution, the median $H_0$ and its 68\% credible interval are given in $\text{km}\,\text{s}^{-1}\,\text{Mpc}^{-1}$ on the left, the specific contribution from the multipole to the error budget calculated as in \eqref{eq_sigma_multipole} expressed as a percentage of the median $H_0$ is displayed in the middle, and that from multipole-free lensing galaxies is displayed on the right. More specifically, the $\sigma_{\text{multipole}}$ and $\sigma_{\text{no multipole}}$ are the mean values of the two sides of the 68\% credible intervals.}
\label{table_final_pdf_h0}
\end{table*}

Because boxy and discy galaxies have opposite impacts on the inferred value of $H_0$, the final distribution is overall unbiased at the subpercent level. This result is independent of the value of the S/N  considered. However, by increasing the fraction of boxy galaxies, we see that the distribution is slightly driven to lower values of $H_0$ by less than 1\%. The larger scatter on $H_0$ when the S/N decreases is mainly due to the larger uncertainty for a fixed value of $a_4$ which is stronger for lower S/N, especially for high multipolar deformations.

Beyond bias, a noise term due to multipole presence is expected. If we consider that a multipole introduces an uncertainty on the measurement of $H_0$ that is independent from the other sources of uncertainty, that is when no multipoles are present in the lensing galaxy, we can calculate the multipole contribution to the total uncertainty budget. Indeed, we have
\begin{equation}
     \sigma_{\text{multipole}}^2 = \sigma_{\text{tot}}^2 - \sigma_{\text{no multipole}}^2,
     \label{eq_sigma_multipole}
\end{equation}
where $\sigma_{\text{tot}}$ is the total uncertainty quoted in Table \ref{table_final_pdf_h0}, $\sigma_{\text{multipole}}$ is the contribution from the multipoles, and $\sigma_{\text{no multipole}}$ is the uncertainty when modelling images with a multipole-free lensing galaxy. 

The $\sigma_{\text{no multipole}}$ is obtained by summing the PDFs of the $a_4=0.00$ case with the different configurations weighted by the occurrence of each lens configuration in the calculation of the total PDFs (in Table~\ref{table_final_pdf_h0} or Fig.~\ref{pdf_final_h0}). As a cross would have an uncertainty larger than a cusp or a fold, it is important to weight the configurations to calculate a multipole-free PDF that mimics the one with multipoles because, for example, for the highest S/N, not all configurations yield a sufficiently good fit. We find, for equal occurrence of boxyness and discyness (right panel of Fig.~\ref{pdf_final_h0} and Table~\ref{table_final_pdf_h0}), that the multipoles introduce around 2\% random uncertainty on the $H_0$ distribution (i.e. $\sigma_{\rm {multipole}} / H_0 = 0.02$) for the highest S/N considered, and above 5\% for the lower S/N. If instead we consider a distribution of multipole values similar to \cite{Hao2006} (left panel of Fig.~\ref{pdf_final_h0}  and Table~\ref{table_final_pdf_h0}), the relative uncertainty on $H_0$ due to multipoles drops below 2\% in the highest S/N case, and is between 4.5\% and 5\% for the two other S/N cases. In the lower S/N cases, multipoles constitute the dominant source of uncertainty on $H_0$ (i.e. $\sigma_{\text{multipole}} > \sigma_{\text{no multipole}}$). 

We note that we did not weight our sample by the occurrence of the different configurations in real lensing systems. By systematically trying to fit the four configurations, that is the two cusps, the fold, and the cross, we aim at encompassing all possibilities. However, in real cases, some configurations might be more likely than others, and those occurrences may also change from one sample to another. We did not take into account this effect but we are confident that the effect is negligible in terms of systematic bias at a population level. However, it may modify the exact value of the uncertainty on $H_0$.

\subsection{Impact for the TDCOSMSO analysis}

The TDCOSMO/H0LiCOW collaboration combined the results of seven lenses to make their $H_0$ inference \citep{HolicowXIII, TDCOSMOIV}. Those seven lenses were imaged in one or more spectral bands, usually including the F160W/WFC3 HST band with exposure times varying between 2000s and 26000s. Those NIR images are \textit{a priori} the most favourable for analysing extended lensed features (see discussion in Sect. \ref{subsect_mock_crea}), which are the most prone to deformation from boxyness or discyness. Their Einstein radii range between 0.8$\arcsec$ and 2$\arcsec$ \citep{TDCOSMOIV}. The host galaxies are sometimes significantly brighter than the quasar (e.g. the contrast in RXJ1131, whose host is brighter than the quasar by 0.9 mag, which is stronger than what we considered in this study, i.e. a 0.5 mag difference) and sometimes the quasar dominates the source flux \citep[e.g. WFI2033 with its quasar being 1.5 mag brighter than the host; ][]{Ding2021_constrast}. Overall, the TDCOSMO systems are generally very bright, with very high-quality data. For most of the TDCOSMO lenses, we expect large-amplitude multipoles to be detectable if present, and thus to have an unbiased $H_0$ inference already at a single-galaxy level. However, a detailed analysis for each lens is needed to assess the exact detectability of multipoles and its potential impact on the $H_0$ inference for a given lens. Doing so is possible by creating a mock of each system with increasing values of $a_4$, as done in Sect. \ref{section_method}, to see the multipole threshold at which patterns in the residuals are detectable. By combining this with an analysis of the light of the lensing galaxy, it is possible to assess a range of plausible values of $a_4$, and to quantify how they may impact $H_0$ inference.

As a population, the seven lenses should yield an unbiased combined inference of $H_0$: their sample is drawn from distributions centred on the true $H_0$ at a subpercent level (see Fig. \ref{pdf_final_h0} and Table \ref{table_final_pdf_h0}). Indeed, considering the TDCOSMO data quality, we conceive that for most lenses, the multipoles are detectable (typically, red or sometimes green curve in Fig. \ref{pdf_final_h0}). Hypothetically, assuming the worst-case scenario ---that the multipole were not ever detectable (yellow curve in Fig. \ref{pdf_final_h0})--- as long as the lenses are typical of the elliptical population (i.e. with balanced quantity of boxy and discy galaxies and with a typical distribution of $a_4$ as in \cite{Hao2006}), the bias on $H_0$ would still be below the percent level (albeit with increased scatter).

We note that in the TDCOSMO sample, the different single-galaxy inferences are combined by multiplying their PDFs to find the underlying true $H_0$ and the error on this measurement. The distributions we present in this work (Fig. \ref{a4vsH0_levels}) are the underlying distributions of $H_0$ from which a sample will be drawn. Thus, the uncertainties reported in Table \ref{table_final_pdf_h0} are the underlying distribution width, not the error one would get by combining $x$ independent $H_0$ inferences from $x$ lenses (the combined inference being increasingly precise as more lenses are included). The uncertainties on the combined $H_0$ inference quoted in the TDCOSMO analysis \citep{HolicowXIII, TDCOSMOIV} already take into account any broadening due to the possible presence of multipoles.

\section{Conclusion}
\label{section_conclu}

State-of-the-art lens modelling generally assumes that the projected mass density of lensing galaxies follows a perfect ellipse or a combination of elliptical baryonic and dark components. However, boxy or discy isophotal profiles, also referred to as fourth-order multipoles, are observed in massive elliptical galaxies. While the amplitude of those multipoles is unknown for the total mass distribution, lensing galaxies are baryon-dominated in their inner regions. It  therefore seems reasonable to assume that the results of studies investigating boxyness and discyness in the light of galaxies are also applicable to the mass of lensing galaxies.

In this study, we investigated how deviations in galaxy morphology from a pure ellipse modify extended lensed images and become detectable when considering full lens image information. For this purpose, we built an ensemble of mock lensed images of a quasar and its host galaxy quadruply imaged due to strong lensing  by a galaxy that is almost a singular isothermal ellipsoid (SIE): to mimic discy or boxy mass density distributions, we perturbed this density profile by adding fourth-order multipoles with amplitudes matching those detected in the light. Image quality typical of the one achieved with the WFC3 camera on board the Hubble Space Telescope was considered. We then fitted the mock images without any multipolar components.

As our base case, we create a mock image of a very bright QSO+host source with a long exposure time similar to what has been achieved for time-delay cosmography lensed systems. The source we consider is a circular Sersic lensed by an SIE with an Einstein radius of 2\arcsec, with moderate ellipticity ($q = 0.8$) and small multipole perturbation $a_4=0.01$. This multipole amplitude is barely detectable by eye, and is typical of what is generally measured in the light profile of elliptical galaxies ($a_4<$0.02 is expected for 70\% of the systems according to \cite{Hao2006}).

The fitting of this base case showed substantial residuals, suggesting that multipoles, when present, leave ubiquitous imprints in extended strongly lensed images. However, we show that various parameters may influence that detection:
\begin{itemize}
    \item When the SIE, the multipole, the source position, and the shear are aligned, the imprint of the multipoles on the ring can be absorbed in the modelling by a combination of ellipticity and shear. 
    \item Rounder lensing galaxies will create images that are more sensitive to the presence of multipoles. 
    \item The freedom allowed in the lens mass model with a varying power-law slope only slightly decreases  the patterns present in the residuals, independently of the family of models used for the input mass profile. 
    \item Allowing for a structured source luminosity by means of shapelets can absorb substantial residuals coming from the multipoles. 
    \item Naturally, the S/N plays a role: the lower the S/N, the more easily patterns in the residuals are hidden. 
    \item Other parameters, such as the Einstein radius and the host--quasar brightness contrast, may decrease(increase) the detectability of multipoles depending on the host--quasar image blending. 
\end{itemize}

We looked at the impact of the multipole on the $H_0$ retrieval when the model residual is compatible with the noise. We see that boxy and discy galaxies do not have the same impact on $H_0$: the boxy galaxies tend to bias $H_0$ low, while the discy galaxies mostly bias $H_0$ high. In both cases, $H_0$ can be biased by several percent for a given system. However, if we consider a sample of lenses with multipolar contributions drawn in a way that is representative of what is observed in the light of galaxies, high multipolar contributions are less probable and the mix of both boxy and discy galaxies tends to reduce the bias to a level of below 1\%. However, we note that a specific selection of lenses may bias the final $H_0$ inference if for example only boxy galaxies are used. The S/N of the images will mainly broaden the final $H_0$ distribution. 

On a lens-by-lens basis, an upper limit on the bias on $H_0$ cannot be given without performing a simulation that thoroughly emulates the specific characteristics of a given lens. For a given modelled lensing system meant to be used for cosmological inference on its own, this can be performed by following the methodology presented here, creating a mock of the system and adding multipoles to it in order to assess the threshold at which multipoles are detectable in the residuals. In addition, one can also look at the lensing galaxy light profile. Multipoles in the light can in principle be used as a proxy to the presence of multipoles in the total mass, with the caveat that the presence of multipoles in the dark matter is unsettled.

However, instead of looking at the bias for each system individually, one can consider a population approach. If using a wide sample of lensing galaxies displaying multipolar distribution representative of what is observed in the light of elliptical galaxies, we show that the combined $H_0$ should not be biased by the presence of multipole structure. This is also true for the TDCOSMO data. Our results suggest that thanks to the high quality of those data, and the ability of the models to reproduce them down to the noise, the bias on the combined $H_0$ inference from the seven lenses of TDCOSMO is expected to be below the 1\%\ level. Furthermore, the uncertainties quoted in published analyses already take into account any slight broadening caused by to multipolar components. 

Beyond the cosmology inference, this study also shows that, with high-S/N data, the multipoles have an imprint on the image of the extended source. Those observational constraints may be instrumental to better quantifying the contribution of multipoles to flux-ratio anomalies in lensed quasars \citep{Mao1998,Kochanek2004Dalal,Gilman2019}. This work might therefore help in extending existing studies of multipolar-like azimuthal perturbation as the source of flux ratio anomalies \citep[e.g.][]{Keeton2003,Xu2015} by calculating the effective contribution of the multipoles to the flux-ratio anomalies based on the perturbations seen in the ring. However, the accuracy with which multipoles can be recovered from the lens models has not yet been tested, and is left for future work.

\begin{acknowledgement}

The authors thank Daniel Gilman, Martin Millon, Veronica Motta, Sampath Mukherjee, Anna Pasquali, Sherry Suyu, Tommaso Treu, Jenny Wagner, Liliya L.R. Williams, and Kenneth Wong for their useful remarks on this work.

This work makes use of lenstronomy v1.5.1 \citep{lenstro2018} and of the following Python packages : Python v3.6.5 \citep{Python1,Python2}, Astropy \citep{astropy:2013,astropy:2018}, Numpy \citep{Numpy}, Scipy \citep{scipy}, Matplotlib \citep{Matplotlib}, Pandas \citep{pandas1,pandas2}, Photutils \citep{photutils}.

This project has received funding from the European Research
Council (ERC) under the European Union’s Horizon
2020 research and innovation programme (COSMICLENS :
grant agreement No 787886). This programme is supported by the Swiss NationalScience Foundation (SNSF). DDX acknowledges Tsinghua University Initiative Scientific Research Program ID 2019Z07L02017 and the Chinese Academy of Sciences (CAS)
project no. 114A11KYSB20170054. GV has received funding from the European Union’s
Horizon 2020 research and innovation programme under the Marie
Sklodovska-Curie grant agreement No 897124.
\end{acknowledgement}

\bibliographystyle{aa} 
\bibliography{biblio}

\begin{appendix}

\section{Source reconstruction with shapelets}
\label{appendix_shap}

\begin{figure*}
    \centering
    \includegraphics[width=0.9\textwidth]{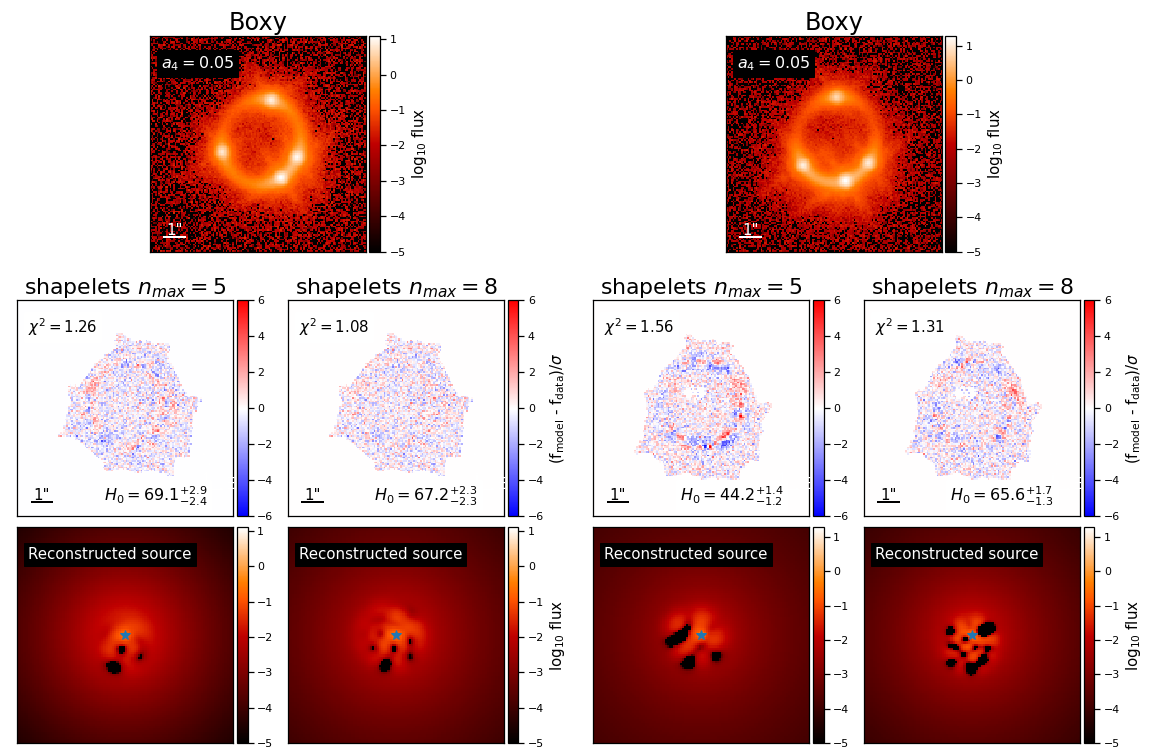}
    \includegraphics[width=0.9\textwidth]{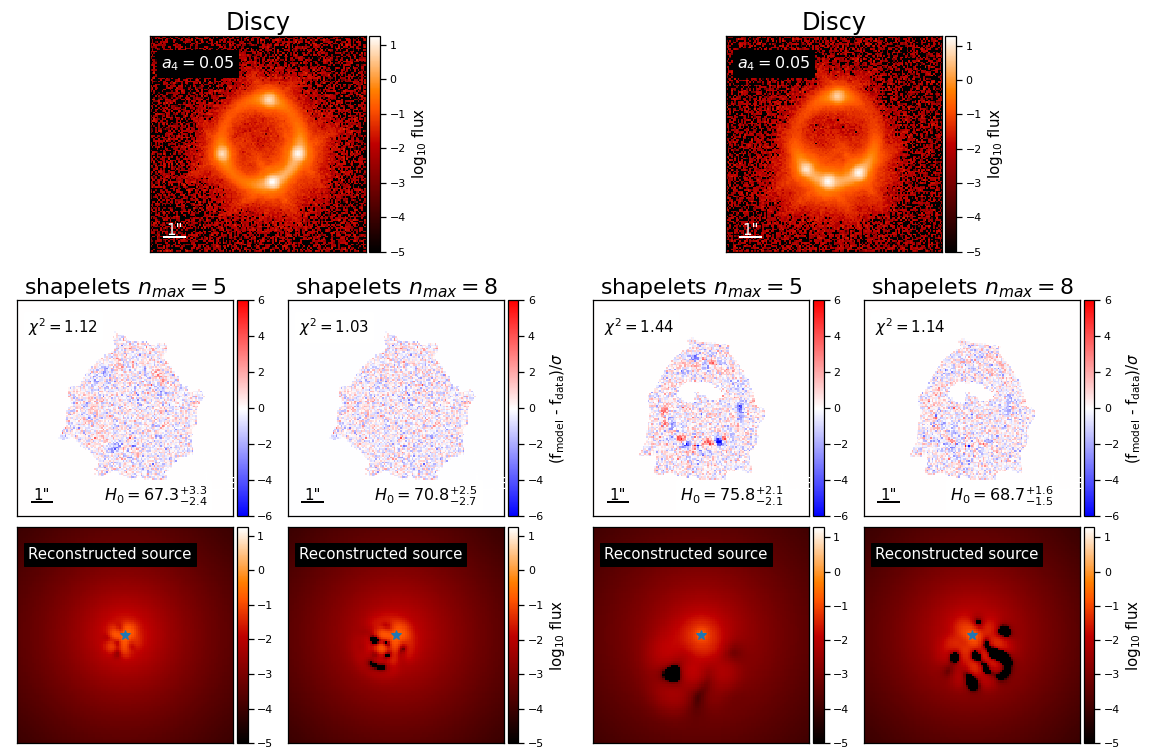}
    \caption{Shapelet reconstruction of the source for four mock systems with $a_4=0.05$, 3000s exposure time and a bright source. The displayed configurations are either \textit{fold} configuration (left) or \textit{cusp\_s} configuration (right). The input lensing galaxy is either boxy (top) or discy (bottom). For each mock system, we consider two fitting setups: including shapelets in the source model with maximal polynomial order $n_{\text{max}} = 5$ or $n_{\text{max}} = 8$. The associated residuals and source shape reconstruction are displayed. For good models (i.e. $\chi^2<1.2$), the morphology of the source gets less realistic as $n_{\text{max}}$ increases.} 
    \label{res_shap8}
\end{figure*}

In Sect. \ref{influ_shap}, we model mock images with additional freedom in the source shape model. We present in Fig.~\ref{res_shap8} the source reconstruction obtained when modelling mock systems created with $a_4 = 0.05$ (last row of Table \ref{big_result_table}) with a source more complex than a smooth circular Sersic. For this purpose, we added shapelets to the source light distribution. We present the results obtained with two different maximal polynomial orders $n_{\text{max}}$ of the shapelets. We see that more complex clumpy structures appear in the source with the increase of $n_{\text{max}}$. Those complex structures create lensed images that mimic those produced by multipoles, sometimes even achieving a reasonably good fit. The change in the source morphology may however hardly mimic structures seen in real galaxies. Models that predict such a source morphology would probably be considered as a symptom of a shortcoming in the macro-model of the lensing galaxy. We suggest that in the future, a machine-learning-based strategy could help to identify such features as artefacts caused by an incorrect macro model
in an objective manner.

\section{Other mass profiles}
\label{appendix_mass_profile}
In Sect. \ref{influ_slope}--\ref{influ_compo}, mass models deviating from the base setup are used. A power-law model is used in the fitting procedure in Sect. \ref{influ_slope}--\ref{influ_compo}. In addition, a composite model made of a baryonic component represented by a Chameleon profile and a dark matter component emulated with a NFW profile are used to create the mock lensing galaxy in Sect. \ref{influ_compo}. Hereafter, we specify the equations associated to those mass profiles.
\subsection{Power-law model}
\label{appendix_pl}
The convergence of an elliptical power-law profile is given by:
\begin{equation}
    \kappa(\theta_1,\theta_2) = \frac{3-\gamma'}{2} \left( \frac{\theta_E}{\sqrt{q\theta_1^2 + \theta_2^2/q}} \right)^{\gamma'-1} 
,\end{equation}
where ($\theta_1,\theta_2$) are the coordinates in a reference frame oriented along the main axis of the power-law ellipsoid, $\gamma'$ is the slope of the profile ($\gamma'=2$ corresponds to the isothermal case), $\theta_E$ is the Einstein radius, and $q$ is the axis ratio.
\subsection{Chameleon model}
\label{appendix_cham}
The Chameleon mass model \citep[][Gomer et al. in prep]{Dutton2011,Suyu2014} is a profile with finite mass equal to the difference between two non-singular isothermal ellipsoids (NIEs) with the same normalisation but different core radii. It also mimics a Sersic profile for Sersic indexes roughly between 1 and 4. This profile is mostly used to represent baryonic matter in mass as well as in light. One formulation of convergence of the Chameleon profile is the following:

\begin{align}
    \kappa(\theta_1,\theta_2) = \frac{L_0}{1+q} \left(\frac{1}{\sqrt{\theta_1^2+\theta_2^2/q^2 + 4 w_c^2 /(1+q)^2 }}\right. -  &  \nonumber \\
    \left. \frac{1}{\sqrt{\theta_1^2+\theta_2^2/q^2 + 4 w_t^2 /(1+q)^2 }} \right),&
\end{align}

where ($\theta_1,\theta_2$) are the coordinates in a reference frame oriented along the main axis of the Chameleon ellipsoid, $L_0$ is a normalisation factor, $q$ is the axis ratio, $w_c$ is a proxy for the core radius of the first NIE, and $w_t$ is a proxy for that of the second NIE. 

\begin{figure}
    \centering
    \includegraphics[width=0.49\textwidth]{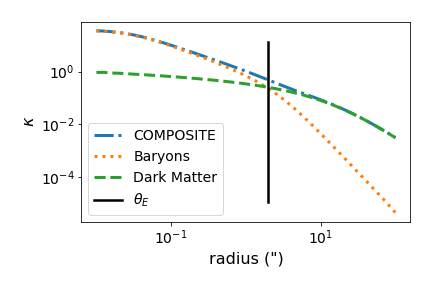}
    \caption{Radial convergence profile of the composite model used in Sect. \ref{influ_compo}. The baryonic component emulated by a Chameleon profile (dotted orange), the dark matter component rendered with a NFW profile (dashed green), and the sum of both components (dash-dotted blue) are represented. The vertical solid black line indicates the Einstein radius of the total composite mass profile.}
    \label{fig_composite_radial_profile}
\end{figure}

In Sect. \ref{influ_compo}, we use the following values: $q=0.8, w_c=0.024, w_t=2.8$, and $L_0$ such that the deflection angle at 1$\arcsec$ (if the Chameleon were circular) is 1.6$\arcsec$. Such a Chameleon profile mimics a Sersic profile of index $n=4$ and effective radius $R_{\rm eff}=2.3\arcsec$. The radial profile of this Chameleon is presented with the dotted orange curve in Fig. \ref{fig_composite_radial_profile}. 

\subsection{NFW model}
\label{appendix_nfw}
The Navarro-Frenk-White \citep{NavarroFrenkWhite1996} mass profile is the most commonly used profile to represent dark matter halos. The three-dimensional density of such a circular profile is given by:
\begin{equation}
    \rho(r)=\frac{\rho_0}{(r/r_s)(1+r/r_s)^2}
,\end{equation}
where $r$ is the radius, $\rho_0$ the central density, and $r_s$ the scale radius. We refer to \cite{GolseKneib2002} for a more complicated formulation of such a profile with ellipticity introduced in the lensing potential, as used in our simulations with \texttt{lenstronomy}.

In Sect. \ref{influ_compo}, we use the following values: $\rho_0=2.3\times10^{15}~\textup{M}_\odot/\textup{Mpc}^3$, $R_s=0.071~\textup{Mpc}$, and the ellipticity is chosen such that the axis ratio in equal to 0.8 in convergence. Physically, this profile has a virial radius $R_{200}=0.42~\textup{Mpc}$ and a virial mass $M_{200} = 1.1\times 10^{13}~\textup{M}_\odot$. The radial profile of this elliptical NFW is drawn in dashed green at Fig. \ref{fig_composite_radial_profile}.

\end{appendix}

\end{document}